\documentclass[12pt]{article}
\usepackage[utf8]{inputenc}

\usepackage[textwidth=6.8in,top=1.18in,bottom=1.18in]{geometry}
\usepackage[margin=1cm]{caption}

\title{\vspace{-4cm} Proximal Inference for Indirect and Intervening Effects in Population Interventions}
\author{
Yang Bai$^1$, Yifan Cui$^2$, and Baoluo Sun$^1$\\
$^1$ Department of Statistics and Data Science, National University of Singapore\\
$^2$ Center for Data Science, Zhejiang University
}
\date{}

\usepackage{amsthm,amsmath,amssymb,bbm,bm}
\usepackage{natbib}
\usepackage{multirow}
\usepackage[pdftex]{graphicx}
\usepackage{wrapfig}
\usepackage{booktabs}
\usepackage{tikz}
\usepackage{centernot}
\usepackage{array}
\usepackage{url}
\usepackage{algorithmic}
\usepackage[linesnumbered, ruled, vlined]{algorithm2e}
\usepackage{mathrsfs}
\usepackage{dsfont}
\usepackage{titling}
\usepackage{relsize}
\usepackage{rotating}
\usepackage{enumitem}
\usepackage{titling}
\usepackage{float}
\usepackage{subcaption}
\usepackage[colorlinks, linkcolor=blue, anchorcolor=blue, citecolor=blue]{hyperref}

\newtheorem{assumption}{Assumption}

\newtheorem{theorem}{Theorem}[section]
\newtheorem{proposition}{Proposition}[section]

\newtheorem{corollary}{Corollary}[section]

\newtheorem{example}{Example}

{\textwidth=6.5in}

\newcommand{\expit}{{\text{expit}}}

\newcommand{\cM}{{\mathcal{M}}}

\newcommand{\cH}{{\mathcal{H}}}
\newcommand{\cG}{{\mathcal{G}}}

\newcommand{\cN}{{\mathcal{N}}}

\newcommand{\indep}{\rotatebox[origin=c]{90}{$\models$}}
\newcommand{\E}{\mathbb{E}}

\begin{document}

\maketitle
\vspace{-1cm}
\begin{abstract}
    Unmeasured confounding, unethical exposure, and ill-defined interventions pose significant challenges to evaluating policy-relevant mediation estimands in medicine and public health. In observational studies involving harmful exposures, the population intervention indirect effect (PIIE) is often more salient than the natural indirect effect, as the latter relies on hypothetical interventions that may be ethically or practically unfeasible. While the PIIE can be identified via the generalized front-door criterion under unmeasured exposure-outcome confounding, existing estimation methods typically assume the absence of unmeasured confounding for the mediator. Furthermore, when the exposure corresponds to ill-defined interventions, the standard PIIE criterion fails; however, the generalized front-door formula may still identify the causal effect of an intervening variable designed to capture the indirect effect. This paper develops a unified identification and estimation framework for the PIIE and the causal effect of an intervening variable in settings with pervasive unmeasured confounding affecting exposure-mediator, exposure-outcome, and mediator-outcome relationships. Specifically, we leverage observed covariates as proxy variables to construct three distinct identification strategies within a proximal causal inference framework. We characterize the semiparametric efficiency bound for the target estimands and develop multiply robust, locally efficient estimators that remain consistent under partial model misspecification. The finite-sample performance of our estimators is demonstrated through simulations. Finally, we apply our methodology to study the indirect effect of alcohol consumption on depression risk as mediated by depersonalization symptoms.
\end{abstract}


\section{Introduction}
\label{s:intro}

Investigating policy-relevant estimands in causal mediation analysis often faces three concurrent challenges: unmeasured confounding, unethical exposures, and ill-defined interventions. While the proximal causal inference framework \citep{miao2018identifying, tchetgen2020introduction, cui2023semiparametric} addresses unmeasured confounding for both the average causal effect (ACE) and the natural indirect effect (NIE) \citep{dukes2023proximal}, it remains limited in complex settings. First, the NIE is ethically untenable for harmful exposure, making the population intervention indirect effect (PIIE) \citep{hubbard2008population} a more appropriate estimand. Second, existing proximal methods require well-defined interventions; applying them to non-intervenable exposures risks a ``replication crisis'' \citep{wen2024causal}, as these variables may not be operationalized consistently across future studies.

Recent advances handle unethical and non-intervenable exposures separately. The PIIE is identifiable via the generalized front-door criterion \citep{fulcher2020robust}, while non-intervenable exposures can be managed using intervening variables to capture separable indirect effects \citep{robins2022interventionist, stensrud2023conditional}. \cite{wen2024causal} unified these by applying the generalized front-door formula to intervening variables. However, a critical limitation persists: these methods only accommodate unmeasured confounding in the exposure-outcome relationship and fail when unobserved factors directly affect the mediator.

In this paper, we propose a unified proximal framework that overcomes these limitations. We establish a more general setup than \cite{fulcher2020robust} and \cite{wen2024causal}, permitting unmeasured confounders to simultaneously affect the exposure, mediator, and outcome. By leveraging proxy variables, we prove that both the PIIE and intervening variable causal effects are nonparametrically identifiable even under this pervasive confounding. Ultimately, this successfully addresses the dual challenges of unethical exposures and non-intervenable variables within a single, robust framework.

\subsection{Prior works on population intervention indirect effect}

\begin{example}\emph{\textbf{(Mediation analysis with unethical exposure)}}
    Alcohol use and major depression are highly comorbid \citep{boden2011alcohol}, potentially mediated by depersonalization symptoms \citep{raimo1999alcohol}. Evaluating this pathway is challenging because pervasive unmeasured confounding $(U)$, such as genetic predisposition or childhood trauma, jointly affects the exposure, mediator, and outcome. Furthermore, because alcohol is harmful, estimating the NIE requires ethically untenable interventions on the exposure level.
  \label{example: alcohol consumption}
\end{example}

To address settings like Example~\ref{example: alcohol consumption}, where standard NIE interventions are unethical because hypothetical interventions that set exposure to harmful levels are unrealistic, \cite{hubbard2008population} proposed the population intervention effect (PIE). \cite{fulcher2020robust} later decomposed the PIE to isolate the PIIE. The PIIE captures the effect of shifting the mediator from the natural value to its potential value under no exposure, while holding the exposure at its naturally observed level.

Traditional PIIE inference assumes all confounders are observed \citep{hubbard2008population}. While \cite{fulcher2020robust} relaxed this using a generalized front-door criterion to handle unmeasured exposure-outcome confounding, their method fails if unmeasured factors directly affect the mediator. Concurrently, proximal causal inference has emerged to mitigate unmeasured confounding using proxy variables \citep{miao2018identifying, tchetgen2020introduction, cui2023semiparametric, ying2023proximal, sverdrup2023proximal, qi2024proximal}. However, mediation extensions of this framework either assume the mediator is hidden but still unconfounded \citep{ghassami2024causal}, or focus strictly on the NIE \citep{dukes2023proximal}. Consequently, inferring the PIIE under pervasive unmeasured confounding, where $U$ simultaneously affects the exposure, mediator, and outcome, remains an unsolved challenge.

\subsection{Prior works on the causal effect of intervening variables}

\begin{example}\emph{\textbf{(Mediation analysis with non-intervenable exposure)}}
      Chronic pain is a leading precursor to opioid abuse and mortality \citep{dowell2016cdc}. While evaluating the indirect effect of chronic pain on mortality via opioid prescription is prevalent \citep{inoue2022causal}, chronic pain itself is notoriously treatment-resistant and difficult to manipulate \citep{wen2024causal}. Furthermore, unmeasured socio-psychological factors frequently confound the complex relationships between pain, opioid prescription, and mortality.
  \label{example: chronic pain}
\end{example}

Because strict interventions on conditions like chronic pain in Example~\ref{example: chronic pain} are infeasible, standard mediation estimands (such as the PIIE or NIE) become conceptually ``ill-defined'' and yield dubious public health implications \citep{holland1986statistics}. To resolve this, \cite{wen2024causal} leveraged separable effects theory \citep{robins2010alternative, robins2022interventionist, stensrud2021generalized, stensrud2023conditional} to focus on an \textit{intervening variable}, a manipulable descendant (e.g., a physician's perception of the patient's pain) of the non-intervenable exposure that precedes the mediator and transmits the exposure's indirect effect.

\cite{wen2024causal} demonstrated that the generalized front-door formula can identify this intervening variable effect under unmeasured exposure-outcome confounding. However, their framework requires the exposure-mediator and mediator-outcome relationships to be entirely unconfounded. In reality, latent factors like a patient's psychological state often confound all three pathways simultaneously. Although proximal methods for conditional separable effects have been proposed  \citep{park2024proximal}, identifying the causal effect of an intervening variable under such pervasive unmeasured confounding, affecting the exposure, mediator, and outcome, remains an open problem.

\subsection{Our contribution}

This article establishes a unified proximal inference framework for mediation analysis when interventions are either unethical (PIIE) or ill-defined (intervening variables). By leveraging proxy variables, we identify these effects even under pervasive unmeasured confounding ($U$) that jointly affect the exposure, mediator, and outcome. Our key contributions are:
\begin{itemize}
    \item \textbf{Nonparametric identification:} We develop three distinct identification strategies using confounding bridge functions. Crucially, these approaches bypass the limitations of the generalized front-door formula by accommodating unmeasured confounding in the exposure-mediator and mediator-outcome relationships.
    \item \textbf{Semiparametric efficiency and multiple robustness:} We derive the efficient influence function and corresponding efficiency bound. Building on this, we propose multiply robust estimators that remain consistent if at least two bridge functions are correctly specified, attaining local efficiency when all models are correct.
    \item \textbf{Debiased machine learning:} Because our estimators satisfy a second-order bias property, they can seamlessly integrate modern machine learning for high-dimensional nuisance parameter estimation. As detailed in the Appendix~\ref{sec: debiased ML}, we demonstrate that cross-fitting preserves $\sqrt{n}$-consistency and asymptotic normality under relaxed convergence rates.
    \item \textbf{Empirical validation:} Through simulations and a clinical application, we quantify the indirect effect of hazardous alcohol consumption on depression, confirming that this risk is significantly elevated via depersonalization symptoms.
\end{itemize}
To our knowledge, this is the first unified framework for PIIE and intervening variable estimation that accounts for pervasive unmeasured confounding across all mediation pathways.

\section{Causal Identification via the Generalized Front-Door Strategy}\label{sec: generalized front-door}

We review two models targeting distinct mediation effects: the population intervention indirect effect \citep{fulcher2020robust} and the causal effect of an intervening variable \citep{wen2024causal}. Both models involve a binary exposure $A \in \left \{ 0, 1 \right \}$, an outcome $Y$, observed covariates $L$, and a mediator $M \in \mathcal{S}$. 

The primary distinction lies in the intervention target: the PIIE assumes well-defined interventions on $A$, whereas the intervening variable framework accommodates ill-defined interventions on $A$ by targeting a modifiable descendant $A_M$. Both frameworks share the same identification requirements: unmeasured confounding is permitted in the $A-Y$ relationship (denoted as $V$ in Figure~\ref{fig: PIIE and intervening variable}), but no unmeasured confounding is allowed for the $A-M$ and $M-Y$ relationships.

\tikzstyle{var} = [circle,very thick,draw=black,fill=white,minimum size=10mm]
\tikzstyle{var_unob} = [circle,very thick,draw=black,fill=gray,minimum size=10mm]
\tikzstyle{var_uninter} = [circle, dashed,very thick,draw=black,fill=white,minimum size=10mm]
\begin{figure}[ht]
    \centering 
    \subfloat[\label{fig: Fulcher}]{
      \begin{tikzpicture}[>=latex, scale=0.7, every node/.style={scale=0.7}]
       \node (A) [var, xshift=-3cm,yshift=0cm]{$\boldsymbol{A}$};
       \node (M) [var, xshift=0cm,yshift=-2cm] {$\boldsymbol{M}$};
       \node (Y) [var, xshift=3cm,yshift=0cm]{$\boldsymbol{Y}$};
       \node (L) [var, xshift=0cm,yshift=2cm]{$\boldsymbol{L}$};
        \node (V) [var_unob, xshift=0cm,yshift=4cm]{$\boldsymbol{V}$};
       \draw[->, very thick] (L) -- (A);
       \draw[->, very thick] (A) -- (Y);
       \draw[->, very thick] (L) -- (M);
       \draw[->, very thick] (A) -- (M);
       \draw[->, very thick] (M) -- (Y);
       \draw[->, very thick] (L) -- (Y);
       \draw[->, very thick] (V) -- (A);
       \draw[->, very thick] (V) -- (Y);
       \draw[->, very thick] (V) -- (L);
       \end{tikzpicture}
    }
    \qquad
    \subfloat[\label{fig: Wen}]{
      \begin{tikzpicture}[>=latex, scale=0.7, every node/.style={scale=0.7}]
        \node (A) [var_uninter, xshift=-3cm,yshift=0cm]{$\boldsymbol{A}$};
       \node (A_M) [var,  xshift=-2cm,yshift=-1.8cm]{$\boldsymbol{A_M}$};
       \node (M) [var, xshift=0cm,yshift=-2cm] {$\boldsymbol{M}$};
       \node (Y) [var, xshift=3cm,yshift=0cm]{$\boldsymbol{Y}$};
       \node (L) [var, xshift=0cm,yshift=2cm]{$\boldsymbol{L}$};
       \node (V) [var_unob, xshift=0cm,yshift=4cm]{$\boldsymbol{V}$};
       \draw[->, very thick] (L) -- (A);
       \draw[->, very thick] (A) -- (Y);
       \draw[->, very thick] (L) -- (M);
       \draw[->, line width = 2.8pt] (A) -- (A_M);
       \draw[->, very thick] (A_M) -- (M);
       \draw[->, very thick] (M) -- (Y);
       \draw[->, very thick] (L) -- (Y);
       \draw[->, very thick] (V) -- (Y);
       \draw[->, very thick] (V) -- (A);
       \draw[->, very thick] (V) -- (L);
       \draw[->, very thick] (V) -- (L);
       \end{tikzpicture}
    }
    \caption{(a) The model from \cite{fulcher2020robust} targets the population intervention indirect effect of $A$ on $Y$ through $M$; (b) The model from \cite{wen2024causal} targets the causal effect of an intervening variable $A_M$. ~ Gray vertices represent unmeasured variables. The intervention on the dashed vertex is ill-defined. The bold arrow represents the deterministic relationship.}
    \label{fig: PIIE and intervening variable}
\end{figure}

\vspace{-0.5cm}
\subsection{Generalized front-door criterion for the PIIE}\label{subsec: generalized frontdoor PIIE}

Following \cite{fulcher2020robust}, we assume interventions on exposure $A$ are well-defined. Let $M(a)$ represent the counterfactual mediator had the exposure been set to $A=a$, and $Y(a,m)$ the potential outcome we would have observed if, possibly contrary to the fact, the exposure and mediator had been assigned to $A=a$ and $M=m$. Based on counterfactual variables $Y(a,m)$ and $M(a)$, we presume $Y(a) := Y(a,M(a))$ and $Y(m) := Y(A,m)$.

When $A=1$ represents a harmful exposure, such as hazardous alcohol consumption in Example~\ref{example: alcohol consumption}, standard natural indirect effects are ethically untenable. Consequently, \cite{hubbard2008population} proposed the population intervention effect:
$\E[Y - Y(0)]$,
which contrasts the observed population outcome with a counterfactual state where no one is exposed to the hazard. \cite{fulcher2020robust} decompose the PIE as follows:
\begin{equation}
    \E\left [ Y - Y(0) \right ] 
    = \underbrace{ \E\left [ Y(A,M(A)) - Y(A,M(0)) \right ] }_{\text{Population Intervention Indirect Effect (PIIE)}} + \underbrace{ \E\left [ Y(A,M(0)) - Y(0,M(0)) \right ] }_{\text{Population Intervention Direct Effect (PIDE)}}. \label{eq_PIE}
\end{equation}
The PIIE captures the indirect effect of $A$ on $Y$ via $M$ by setting the mediator to its counterfactual level under $A=0$ on the mediator, while the exposure remains at its natural level. This type of indirect effect describes the extent to which an intermediate variable mediates the effect of exposure under an intervention that holds the component of exposure directly affecting the outcome at its observed level. Since $\E(Y)=\E\left [Y(A,M(A))\right ]$ is identifiable and $\E\left [ Y(0) \right ] = \E\left [ Y(0, M(0)) \right ]$ can be estimated via proximal inference \cite{cui2023semiparametric}, the primary estimand of interest is the mediation functional $\psi = \E\left [ Y(A, M(0)) \right ]$.

Let $(U, V)$ represent unmeasured, potentially vector-valued confounders, where $U$ is a common cause of $\left \{ A, M, Y \right \}$, and $V$ is a common cause of $\left \{ A, Y \right \}$. To identify $\psi$ using measured covariates $L$, \cite{fulcher2020robust} invoke the \textit{generalized front-door criterion}: $\text{(M1)}~ Y(a,m) = Y ~\text{if}~ A=a ~\text{and}~ M=m.
~\text{(M2)}~ M(a) \indep A | L \quad \forall a \in \left \{ 0, 1 \right \}. 
~\text{(M3)}~ Y(a,m) \indep M(a^{\ast}) | A=a, L \quad \forall a, a^{\ast} \in \left \{ 0, 1 \right \}, m \in \mathcal{S}.$

M1 claims that interventions on the exposure $A$ are well-defined. M2 asserts the absence of the discernible single-world confounding between $M$ and $A$ conditional on $L$. Notably, M3 is a cross-world exchangeability assumption that is untestable even in principle \citep{dawid2000causal, robins2010alternative}, and stronger than only assuming that the $M-Y$ relation does not have unmeasured confounding, which posits the absence of the indiscernible cross-world confounding between $Y$ and $M$ given $A$ and $L$. Figure~\ref{fig: Fulcher} illustrates conditions M1-M3 under the Nonparametric Structural Equation Model with Independent Errors (NPSEM-IE) interpretation of the causal diagram \citep{robins1986new, pearl2009causality}. Importantly, this model permits an unmeasured $A-Y$ confounder ($V$) but assumes no unmeasured confounders ($U$) affect the $A-M$ or $M-Y$ relationships. Under M1-M3, $\psi$ is identified by the \textit{generalized front-door formula}:
\begin{equation}
    \psi = \iiint \E\left [ Y|A=a,M=m,L=l \right ] \mathrm{d}F(m|A=0,l) \mathrm{d}F(a|l) \mathrm{d}F(l).
\label{eq: PIIE-front door}
\end{equation}
While named as a generalization of \textit{Pearl's front-door formula} \citep{pearl2009causality}, \cite{wen2024causal} note that M3 is an untestable cross-world independence assumption that is not required for the original front-door formula to identify the average causal effect.

\subsection{Generalized front-door formula for causal effect of intervening variable}\label{subsec: generalized frontdoor intervening variable}

When interventions on exposure $A$ are ill-defined, such as chronic pain in Example~\ref{example: chronic pain}, standard mediation estimands like the PIIE or NIE risk clinical irrelevance and contribute to the replication crisis in public health \citep{wen2024causal}. Following the theory of separable effects \citep{robins2010alternative, robins2022interventionist, stensrud2021generalized, stensrud2023conditional}, \cite{wen2024causal} propose targeting a modifiable intervening variable $A_M$. In observed data, $A_M$ is deterministically equal to $A$, but it is conceptually distinct as it fully captures the causal pathway from $A$ to $Y$ through $M$. For example, while one cannot strictly intervene on a patient's chronic pain level, one can intervene on a clinician's perception of that pain, which in turn determines opioid prescriptions. A detailed discussion about \cite{wen2024causal} and Example~\ref{example: chronic pain} is presented in the Appendix~\ref{app: intervening variable}.

The counterfactuals induced by interventions on $A_M$ can also be defined analogously. Let $M(a_M)$ and $Y(a_M)$ denote the counterfactual mediator and potential outcome, respectively, had the intervening variable been set to $A_M = a_M$. The intervening variable estimand is defined as the expected counterfactual outcome $\E\left[ Y(a_M) \right]$, for $a_M \in \left \{ 0, 1 \right \}$. As such, the causal effect of an intervening variable $A_M$ is defined as a contrast between $\E\left[ Y(a_M=0) \right]$ and another causal estimand such as $\E\left[ Y(a_M=1) \right]$ or $\E(Y)$. In Example~\ref{example: chronic pain}, note that chronic pain status should no longer be used to determine opioid prescription according to the current guidelines by the Centers for Disease Control and Prevention (CDC) \citep{dowell2016cdc,inoue2022causal}, then there is interest in evaluating this modified prescription policy, by comparing the natural perception on opioid prescription with the suggested policy that the doctor disregards a patient's chronic pain in their decisions on opioid prescription. Hence, $\E\left[ Y - Y(a_M=0) \right]$ should be of primary interest in this example. 

With a slight abuse of notation from \cite{stensrud2021generalized}, let the superscript `G' refer to a future trial where $A_M$ is randomly assigned. To identify the intervening variable estimand, \cite{wen2024causal} impose the following conditions:
$\text{(N1)}~ \text{If}~ A_M=a_M, \text{then}~ M(a_M) = M;
~\text{If}~ A_M=a_M, \text{then}~ Y(a_M) = Y;
\text{(N2)}~ A = A_M.
\text{(N3)}~ Y^{G} \indep A_M^{G} | A^{G}, M^{G}, L^{G}; M^{G} \indep A^{G} | A_M^{G}, L^{G}.$ 

N1 states that interventions on the intervening variable $A_M$ are well-defined, which accommodates interventions on the exposure $A$ to be ill-defined. N2 implies that $A_M$ is deterministically equal to $A$. N3 is the so-called dismissible component condition assumed in the scenario of `G', substantially ensuring the absence of unmeasured confounding in the $A-M$ and $M-Y$ relationships while permitting an unmeasured common cause $V$ of $A$ and $Y$. Additionally, N3 would fail if there is a direct effect of $A_M$ on $Y$ not mediated by $M$, or $A$ exerts an effect on $M$ not through $A_M$. Figure~\ref{fig: Wen} provides a graphical illustration of conditions N1-N3 under the NPSEM-IE interpretation. Under N1-N3, the intervening variable estimand $\E\left[ Y(a_M) \right]$ is identified by the generalized front-door formula:
\begin{equation}
    \E\left[ Y(a_M) \right] = \iiint \E\left [ Y|A=a,M=m,L=l \right ] \mathrm{d}F(m|A=a_M,l) \mathrm{d}F(a|l) \mathrm{d}F(l).
  \label{eq: intervening-front door}
\end{equation}
Crucially, while the functional form matches Equation~\eqref{eq: PIIE-front door}, the identifying assumptions are fundamentally different. The intervening variable framework bypasses the untestable cross-world exchangeability M3 required for the PIIE, providing a more robust and interpretable alternative when exposure interventions are ambiguous. However, like the PIIE framework, it remains unable to handle pervasive confounding $U$ that affects $A$, $M$, and $Y$ simultaneously.

\section{Identification Methods under a General Confounding Mechanism}\label{sec: identification}

\subsection{Proxy variables and confounding bridge functions}

While the generalized front-door formulas \eqref{eq: PIIE-front door} and \eqref{eq: intervening-front door} identify mediation effects, they rely on the absence of unmeasured $A-M$ and $M-Y$ confounding. In practice, unmeasured factors $U$ (e.g., latent mental complications in Example~\ref{example: alcohol consumption} or clinician disposition in Example~\ref{example: chronic pain}) often simultaneously affect the exposure, mediator, and outcome. Under this general unmeasured confounding mechanism (see Figure~\ref{fig: proximal PIIE and intervening}), the conditional exchangeability M2, the cross-world independence M3, and the dismissible component condition N3 are violated, and the generalized front-door formula no longer holds.

\tikzstyle{var} = [circle, very thick,draw=black,fill=white,minimum size=10mm]
\tikzstyle{var_unob} = [circle, very thick,draw=black,fill=gray,minimum size=10mm]
\tikzstyle{var_uninter} = [circle, dashed,very thick,draw=black,fill=white,minimum size=10mm]
\begin{figure}[ht]
    \centering 
    \subfloat[\label{fig: proximal PIIE}]{
       \begin{tikzpicture}[>=latex, scale=0.7, every node/.style={scale=0.7}]
       \node (A) [var, xshift=-3cm,yshift=0cm]{$\boldsymbol{A}$};
       \node (M) [var, xshift=0cm,yshift=-2cm] {$\boldsymbol{M}$};
       \node (Y) [var, xshift=3cm,yshift=0cm]{$\boldsymbol{Y}$};
       \node (X) [var, xshift=0cm,yshift=2cm]{$\boldsymbol{X}$};
       \node (Z) [var, xshift=-2.5cm,yshift=3cm]
       {$\boldsymbol{Z}$};
       \node (W) [var, xshift=2.5cm,yshift=3cm]
       {$\boldsymbol{W}$};
       \node (U) [var_unob, xshift=0cm,yshift=4cm]{$\boldsymbol{U}$};
       \node (V) [var_unob, xshift=0cm,yshift=6cm]{$\boldsymbol{V}$};
       \draw[->, very thick] (X) -- (A);
       \draw[->, very thick] (A) -- (Y);
       \draw[->, very thick] (X) -- (M);
       \draw[->, very thick] (X) -- (Z);
       \draw[->, very thick] (X) -- (W);
       \draw[->, very thick] (Z) -- (A);
       \draw[->, very thick] (W) -- (Y);
       \draw[->, very thick] (A) -- (M);
       \draw[->, very thick] (M) -- (Y);
       \draw[->, very thick] (X) -- (Y);
       \draw[->, very thick] (U) -- (Y);
       \draw[->, very thick] (U) -- (A);
       \draw[->, very thick] (U) -- (X);
       \draw[->, very thick] (U) -- (X);
       \draw[->, very thick] (U) -- (Z);
       \draw[->, very thick] (U) -- (W);
       \draw[->, very thick] (V) -- (U);
       \draw[->, very thick] (V) -- (A);
       \draw[->, very thick] (V) -- (Y);
       \draw[<-, very thick] (0.3,-1.6) arc (285:430 : 1 and 2.8);
       \end{tikzpicture}
    }
    \qquad
    \subfloat[\label{fig: proximal intervening variable}]{
    \begin{tikzpicture}[>=latex, scale=0.7, every node/.style={scale=0.7}]
       \node (A) [var_uninter, xshift=-3cm,yshift=0cm]{$\boldsymbol{A}$};
       \node (A_M) [var,  xshift=-2cm,yshift=-1.8cm]{$\boldsymbol{A_M}$};
       \node (M) [var, xshift=0cm,yshift=-2cm] {$\boldsymbol{M}$};
       \node (Y) [var, xshift=3cm,yshift=0cm]{$\boldsymbol{Y}$};
       \node (X) [var, xshift=0cm,yshift=2cm]{$\boldsymbol{X}$};
       \node (Z) [var, xshift=-2.5cm,yshift=3cm]
       {$\boldsymbol{Z}$};
       \node (W) [var, xshift=2.5cm,yshift=3cm]
       {$\boldsymbol{W}$};
       \node (U) [var_unob, xshift=0cm,yshift=4cm]{$\boldsymbol{U}$};
       \node (V) [var_unob, xshift=0cm,yshift=6cm]{$\boldsymbol{V}$};
       \draw[->, line width = 2.8pt] (A) -- (A_M);
       \draw[->, very thick] (A_M) -- (M);
       \draw[->, very thick] (X) -- (A);
       \draw[->, very thick] (A) -- (Y);
       \draw[->, very thick] (X) -- (M);
       \draw[->, very thick] (X) -- (Z);
       \draw[->, very thick] (X) -- (W);
       \draw[->, very thick] (Z) -- (A);
       \draw[->, very thick] (W) -- (Y);
       \draw[->, very thick] (M) -- (Y);
       \draw[->, very thick] (X) -- (Y);
       \draw[->, very thick] (U) -- (Y);
       \draw[->, very thick] (U) -- (A);
       \draw[->, very thick] (U) -- (X);
       \draw[->, very thick] (U) -- (X);
       \draw[->, very thick] (U) -- (Z);
       \draw[->, very thick] (U) -- (W);
       \draw[->, very thick] (V) -- (U);
       \draw[->, very thick] (V) -- (A);
       \draw[->, very thick] (V) -- (Y);
       \draw[<-, very thick] (0.3,-1.6) arc (285:430 : 1 and 2.8);
       \end{tikzpicture}
    }
    \caption{General confounding mechanism: $U$ is a common cause of $\left \{ A, M, Y \right \}$; $V$ is a common cause of $\left \{ A, Y \right \}$. ~ (a) The proximal model targets the population intervention indirect effect; (b) The proximal model targets the causal effect of an intervening variable $A_M$. ~ Proxy variables $Z$ and $W$ are sufficient to account only for the confounding arising from $U$. Gray vertices represent unmeasured variables. The intervention on the dashed vertex is ill-defined. The bold arrow represents the deterministic relationship.}
    \label{fig: proximal PIIE and intervening}
\end{figure}

To mitigate this bias, we postulate that a subset of observed covariates $L = (X, Z, W)$ can serve as proxy variables for the unmeasured confounder $U$. Here, $X$ denotes baseline covariates; $Z$ represents exposure-inducing confounding proxies (potential causes of $A$, related to $M$ and $Y$ only through $A$, $X$, and $U$); and $W$ represents outcome-inducing confounding proxies $W$ (potential causes of $Y$, related to $A$, $M$, and $W$ only through $X$ and $U$).

\begin{assumption}\emph{(Proxy variables)}: 
    (i) $Z \indep \left \{ W,M,Y \right \} \mid U,A,X$; 
    (ii) $W \indep \left \{ A,M \right \} \mid U,X$.
  \label{asm: proxies}
\end{assumption}

As a generalization of the two causal models reviewed in Section~\ref{sec: generalized front-door} to more realistic confounding mechanisms, Figure~\ref{fig: proximal PIIE and intervening} provides graphical illustrations of Assumption~\ref{asm: proxies}. Assumption~\ref{asm: proxies} formalizes the use of $Z$ and $W$ to account for $U$. Note that $Z$ and $W$ are not required to account for $V$, the unmeasured $A-Y$ confounder. Specifically, $A$ and $M$ have no direct effect on $W$, and $Z$ does not directly affect $M$ or $Y$. In addition, $Z$ and $W$ are assumed to be related only through their common measured and unmeasured causes $X$ and $U$. To identify effects using these proxies, we first require a completeness condition, ensuring $Z$ possesses sufficient variability to capture the information in $U$.

\begin{assumption}\emph{(Completeness-$Z$)}:
    (i) For any square-integrable function $g(u)$ and fixed $\left \{ a,m,x \right \}$, if $\E \left [ g(U)|Z,A=a,M=m,X=x \right ]=0$ almost surely, then $g(U)=0$ almost surely.
    (ii) For any square-integrable function $g(u)$ and fixed $x$, if $\E \left [ g(U)|Z,A=0,X=x \right ]=0$ almost surely, then $g(U)=0$ almost surely.
  \label{asm: completeness}
\end{assumption} 

Assumption~\ref{asm: completeness} ensures that any variation in $U$ is fully captured by the corresponding variation in $Z$ when $A$, $M$, and $X$ or $A=0$ and $X$ are fixed. Further discussion of completeness conditions is provided in the Appendix~\ref{app: completeness}. Leveraging an informative $Z$ allows for the existence of outcome confounding bridge functions, which map the observed outcome expectations to the proxies.

\begin{assumption}\emph{(Outcome confounding bridge functions)}:
    Suppose that
    
    (i) there exists outcome confounding bridge function $h_1(w, m, a, x)$ that satisfies
    \begin{align}
        \E(Y|Z,A,M,X) = \E\left[ h_1(W,M,A,X) \mid Z,A,M,X \right]; \label{eq: h1}
    \end{align}
    
    (ii) there exists outcome confounding bridge function $h_0(w, a, x)$, satisfying for $a \in \left \{ 0, 1 \right \}$, 
    \begin{align}
        \E\left [ h_1(W,M,a,X)|Z,A=0,X \right ] = \E\left[ h_0(W,a,X) \mid Z,A=0,X \right]. \label{eq: h0}
    \end{align}
  \label{asm: outcome bridge functions}
\end{assumption}

\vspace{-0.8cm}
Equations~\eqref{eq: h1} and \eqref{eq: h0} are Fredholm integral equations of the first kind, and their solutions are not required to be unique. For conditions that guarantee these equations to admit solutions, we refer readers to \cite{miao2018identifying} and \cite{cui2023semiparametric}.

As a counterpart of Assumption~\ref{asm: completeness}-\ref{asm: outcome bridge functions}, one may leverage $W$ under another completeness assumption to posit the existence of exposure confounding bridge functions $q_0$ and $q_1$.

\begin{assumption}\emph{(Completeness-$W$)}:
     (i) For any square-integrable function $g(u)$ and fixed $\left \{ m,x \right \}$, if $\E[g(U)|W,A=1,M=m,X=x]=0$ almost surely, then $g(U)=0$ almost surely.
     (ii) For any square-integrable function $g(u)$ and fixed $x$, if $\E[g(U)|W,A=0,X=x]=0$ almost surely, then $g(U)=0$ almost surely.
\label{asm: completeness-q}
\end{assumption}

\begin{assumption}\emph{(Exposure confounding bridge functions)}: Suppose that

    (i) there exists the exposure confounding bridge function $q_0(z,x)$ that satisfies
    \begin{align}
        \frac{f(A=1|W,X)}{f(A=0|W,X)} = \E\left [ q_0(Z,X)|W,A=0,X \right ]; \label{eq: q0}
    \end{align}

    (ii) there exists the exposure confounding bridge function $q_1(z,m,x)$ that satisfies
    \begin{align}
        \E\left[ q_0(Z,X)|W,A=0,M,X \right] \frac{f(A=0|W,M,X)}{f(A=1|W,M,X)} = \E\left[ q_1(Z,M,X)|W,A=1,M,X \right]. \label{eq: q1}
    \end{align}
    
  \label{asm: exposure bridge functions}
\end{assumption}

The following proposition provides the conditional moment equations necessary to construct estimators for these exposure confounding bridge functions.

\begin{proposition}
    (i) The function $q_0(z, x)$ satisfies Equation~\eqref{eq: q0} if and only if it solves
    \begin{equation}
        \E\left[ (1-A) q_0(Z,X) -A |W, X \right] = 0. \label{eq: q0_integral}
    \end{equation}
    
    (ii) The function $q_1(Z, M, X)$ satisfies Equation~\eqref{eq: q1} if and only if it solves
    \begin{equation}
        \E\left [ A q_1(Z,M,X) - (1-A) q_0(Z,X) |W,M,X \right ] = 0. \label{eq: q1_integral}
    \end{equation}
    
  \label{propo_integral}
\end{proposition}

\subsection{Proximal identification methods for population intervention indirect effect}\label{sec: proximal PIIE identification}

In this section, we leverage proxy variables to identify the PIIE under the general confounding mechanism illustrated by Figure~\ref{fig: proximal PIIE}. Because the unmeasured common cause $U$ of $\left \{ A, M, Y \right \}$ violates conditional exchangeability M2 and cross-world independence M3, the standard generalized front-door criterion fails. We formalize this latent confounding scenario with the following assumptions. 

\begin{assumption}\emph{(Positivity)}:
    For $a=0,1$ and each $m \in \mathcal{S}$, we have $f_{M|A,U,X}(m|A,U,X)>0$ and $\mathbb{P}(A=a|U,X)>0$ almost surely.
\label{asm: positivity}
\end{assumption}

\begin{assumption}\emph{(Consistency)}:
    (i) $M(a)=M$ almost surely if $A=a$; 
    (ii) $Y(a,m)=Y$ almost surely if $A=a$ and $M=m$.
\label{asm: consistency}
\end{assumption}

\begin{assumption}\emph{(Latent conditional exchangeability)}:
        $M(a) \indep A|U,X$ for $a=0,1$.
\label{asm: latent exchangeability}
\end{assumption}

\begin{assumption}\emph{(Latent cross-world independence)}:
    $Y(a,m) \indep M(a^{\ast})|A=a, U, X$ for $a,a^{\ast}=0,1$ and each $m \in \mathcal{S}$.
\label{asm: latent cross world}
\end{assumption}

Assumption~\ref{asm: positivity} extends the standard positivity condition to include the unmeasured confounder $U$. Assumption~\ref{asm: consistency} states that interventions on exposure $A$ are well-defined. Assumptions~\ref{asm: latent exchangeability}-\ref{asm: latent cross world} relax the generalized front-door conditions of \citep{fulcher2020robust} by assuming $U$ sufficiently accounts for all unmeasured confounding in the $A-M$ and $M-Y$ relationships, while permitting partial unmeasured $A-Y$ confounding via $V$. Figure~\ref{fig: proximal PIIE} encodes a directed acyclic graph that demonstrates Assumptions~\ref{asm: consistency}-\ref{asm: latent cross world} under the NPSEM-IE interpretation of the causal diagram. Analogous to \eqref{eq: PIIE-front door}, we note that under Assumptions~\ref{asm: positivity}-\ref{asm: latent cross world}, the PIIE functional $\psi$ can be expressed by the following \textit{oracle generalized front-door functional}:
\begin{equation}
    \psi = \iiiint \E\left [ Y|u,A=a,m,x \right ] \mathrm{d}F(m|u,A=0,x) \mathrm{d}F(a|u,x) \mathrm{d}F(u|x)\mathrm{d}F(x),
\label{eq: PIIE-front door oracle}
\end{equation}
Because $U$ is unobserved, we propose three distinct strategies to identify $\psi$ using the proxy variables $Z$ and $W$ under the general confounding mechanism illustrated by Figure~\ref{fig: proximal PIIE}. Proofs for all subsequent results are provided in the Supplementary Material.

The proximal outcome regression (P-OR) strategy leverages the informative exposure-inducing proxy $Z$ equipped with Assumption~\ref{asm: completeness}, alongside two nested outcome confounding bridge functions.

\begin{theorem}\emph{(\textbf{P-OR}):}

    Suppose that there exist outcome confounding bridge functions $h_1(W, M, A, X)$ and $h_0(W, A, X)$ that satisfy Assumption~\ref{asm: outcome bridge functions}. Then, under Assumptions~\ref{asm: proxies}, \ref{asm: completeness}, and \ref{asm: positivity}-\ref{asm: latent cross world}, the mediation functional
    $\psi = \E\left[Y(A, M(0))\right]$ is identified by
    \begin{equation}
        \psi = \E\left[ h_0(W,A,X) \right].
    \label{eq: PIIE id}
    \end{equation}
    
 \label{thm: proximal mediation}
\end{theorem}

We refer to Equation~\eqref{eq: PIIE id} as the \textit{proximal population intervention mediation formula}. Unlike the NIE formula in \cite{dukes2023proximal}, which fixes exposure at the treatment level $A=1$, our estimand $\psi$ incorporates a population-level expectation over $A$. This occurs because our outcome confounding bridge functions $h_1$ and $h_0$ account for the observed variation $A$, consistent with the definition of the PIIE estimand $\psi$: a counterfactual where $A$ remains at its observed value while $M$ is set to its value under $A=0$. 

The proximal population intervention mediation formula~\eqref{eq: PIIE id} extends the generalized front-door formula \citep{fulcher2020robust} to settings with unmeasured $A-M$ and $M-Y$ confounding. Similar to the proximal g-formula \citep{tchetgen2020introduction}, identification is achieved through nested bridge functions.  Notably, while our approach identifies indirect effects without additional proxies beyond those used for average causal effects \citep{miao2018identifying, cui2023semiparametric}, it requires a stricter exclusion restriction: proxies $Z$ and $W$ must not have a direct causal effect on the mediator $M$.

Next, we establish the proximal inverse probability weighting (P-IPW) strategy. This approach leverages the outcome-inducing proxy $W$ under the completeness Assumption~\ref{asm: completeness-q} and uses exposure confounding bridge functions $q_0$ and $q_1$.

\begin{theorem}\emph{(\textbf{P-IPW}):}

    Suppose there exist exposure confounding bridge functions $q_0(Z, X)$ and $q_1(Z, M, X)$ satisfying Assumption~\ref{asm: exposure bridge functions}. Under Assumptions~\ref{asm: proxies}, \ref{asm: completeness-q}, and \ref{asm: positivity}-\ref{asm: latent cross world}, the mediation functional $\psi = \E\left[Y(A, M(0))\right]$ is identified by
    \begin{equation}
      \psi = \E\left [ I(A=1) Y q_1(Z,M,X) + I(A=0) Y \right ].
    \label{eq: PIIE id-q1}
    \end{equation} 
    
  \label{thm: proximal mediation-q}
\end{theorem}

While P-OR and P-IPW rely on nested pairs of a single type of bridge function, we propose a proximal hybrid (P-hybrid) strategy. This method combines an outcome confounding bridge function $h_1$ with an exposure confounding bridge function $q_0$, offering a flexible middle ground for identification.

\begin{theorem}\emph{(\textbf{P-hybrid}):}

    Suppose there exists an outcome confounding bridge function $h_1(W, M, A, X)$ satisfying part (i) of Assumption~\ref{asm: outcome bridge functions}, as well as an exposure confounding bridge function $q_0(Z, X)$ satisfying part (i) of Assumption~\ref{asm: exposure bridge functions}. Under Assumptions~\ref{asm: proxies}, part (i) of \ref{asm: completeness}, part (i) of \ref{asm: completeness-q}, and \ref{asm: positivity}-\ref{asm: latent cross world}, the mediation functional $\psi = \E\left[Y(A, M(0))\right]$ is identified by
    \begin{equation}
        \psi = \E\left [ I(A=0) q_0(Z,X) h_1(W,M,1,X) + I(A=0) h_1(W,M,0,X)\right ].
    \label{eq: PIIE id-q0}
    \end{equation}
 
  \label{thm: proximal mediation-h&q}
\end{theorem}

In summary, we establish three complementary strategies to identify the PIIE functional $\psi$, each leveraging a distinct configuration of confounding bridge functions: 
\begin{itemize}
    \item \textbf{P-OR} (Theorem~\ref{thm: proximal mediation}): Employs nested outcome confounding bridge functions $h_1$ and $h_0$.
    \item \textbf{P-IPW} (Theorem~\ref{thm: proximal mediation-q}) Employs nested exposure confounding bridge functions $q_0$ and $q_1$.
    \item \textbf{P-hybrid} (Theorem~\ref{thm: proximal mediation-h&q}): Combines an outcome bridge $h_1$ with an exposure bridge $q_0$.
\end{itemize}
These complementary approaches provide a robust toolkit for causal mediation analysis. By offering multiple identification paths, this proximal framework ensures the PIIE remains identifiable even when standard generalized front-door criteria are violated.

\subsection{Proximal identification methods for intervening variable causal effects}

This section addresses settings where two challenges coexist: general unmeasured confounding and ill-defined exposure interventions. Figure~\ref{fig: proximal intervening variable} provides a graphical demonstration of this model. We focus on the causal effect of an intervening variable $A_M$ (e.g., a doctor's perception of chronic pain in Example~\ref{example: chronic pain}) when direct intervention on the exposure $A$ (e.g., chronic pain) is problematic, to explore the indirect effect of $A$ on $Y$ through $M$. In this scenario, the unmeasured common cause $U$ of $\left \{ A, M, Y \right \}$ violates the dismissible component condition N3, rendering the standard generalized front-door formula~\eqref{eq: intervening-front door} invalid. 

We formalize this model with the following assumptions:

\begin{assumption}\emph{(Intervening consistency)}:
    (i) $M(a_{M}) = M$ almost surely if $A_M = a_M$;
    (ii) $Y(a_{M}) = Y$ almost surely if $A_M = a_M$.
  \label{asm: intervening consistency}
\end{assumption}

\begin{assumption}\emph{(Intervening variable determinism)}:
    $A = A_{M}$ almost surely.
  \label{asm: intervening determinism}
\end{assumption}

\begin{assumption}\emph{(Latent dismissible component conditions)}:
    (i) $Y^{G} \indep A_M^{G} |U^{G}, A^{G}, M^{G}, X^{G}$;
    (ii) $M^{G} \indep A^{G} |U^{G}, A_M^{G}, X^{G}$.
  \label{asm: latent dismissible}
\end{assumption}

Assumption~\ref{asm: intervening consistency} ensures $A_M$ is intervenable, while $A$ might not be. Assumption~\ref{asm: intervening determinism} implies that $A_M$ is always almost equal to $A$. Importantly, Assumption~\ref{asm: latent dismissible} relaxes the standard dismissible component condition N3 in the scenario of `G' considered by Section~\ref{subsec: generalized frontdoor intervening variable}, by accommodating an unmeasured confounder $U$ that has direct effects on $A$, $M$, and $Y$. This assumption implies exclusion restrictions: $A_M$ affects $Y$ only through $M$, and $A$ affects $M$ only through $A_M$. Under these assumptions and positivity, the intervening variable estimand follows the oracle generalized front-door functional:
\begin{equation}
    \E\left[ Y(a_M) \right] = \iiiint \E\left [ Y|u,A=a,m,x \right ] \mathrm{d}F(m|u,A=a_M,x) \mathrm{d}F(a|u,x) \mathrm{d}F(u|x) \mathrm{d}F(x)
  \label{eq: intervening-front door oracle}
\end{equation}

Leveraging proxies $Z$ and $W$ for $U$ as formalized in Assumption~\ref{asm: proxies} and illustrated by Figure~\ref{fig: proximal intervening variable}, we establish three strategies to identify the intervening variable estimand.

\begin{theorem}
    Under Assumptions~\ref{asm: proxies}, \ref{asm: positivity} and \ref{asm: intervening consistency}-\ref{asm: latent dismissible}, we have the following identification results. 

    \emph{(Part 1) (\textbf{P-OR}):} 
     
     Suppose that there exist outcome confounding bridge functions $h_1(W, M, A, X)$ and $h_0(W, A, X)$ that satisfy Assumption~\ref{asm: outcome bridge functions}. Then, under Assumption~\ref{asm: completeness}, the expected counterfactual outcome under an intervention on $A_M$ is identified by
    \begin{equation}
        \E\left[ Y(a_M = 0) \right] = \E\left[ h_0(W,A,X) \right].
    \label{eq: intervening id}
    \end{equation}
     
    \emph{(Part 2) (\textbf{P-IPW}):}

    Suppose that there exist exposure confounding bridge functions $q_0(Z, X)$ and $q_1(Z, M, X)$ that satisfy Assumption~\ref{asm: exposure bridge functions}. Then, under Assumption~\ref{asm: completeness-q}, the expected counterfactual outcome under an intervention on $A_M$ is identified by
    \begin{equation}
      \E\left[ Y(a_M = 0) \right] = \E\left [ I(A=1) Y q_1(Z,M,X) + I(A=0) Y \right ].
    \label{eq: intervening id-q1}
    \end{equation}

    \emph{(Part 3) (\textbf{P-hybrid}):}
    
    Suppose that there exists an outcome confounding bridge function $h_1(W, M, A, X)$ satisfying part (i) of Assumption~\ref{asm: outcome bridge functions}, and an exposure confounding bridge function $q_0(Z, X)$ satisfying part (i) of Assumption~\ref{asm: exposure bridge functions}. Then, under part (i) of Assumption~\ref{asm: completeness} and part (i) of Assumption~\ref{asm: completeness-q}, the expected counterfactual outcome under an intervention on $A_M$ is identified by
    \begin{equation}
        \E\left[ Y(a_M = 0) \right] = \E\left [ I(A=0) q_0(Z,X) h_1(W,M,1,X) + I(A=0) h_1(W,M,0,X)\right ].
    \label{eq: intervening id-q0}
    \end{equation}
    
  \label{thm: intervening id}
\end{theorem}

Theorem~\ref{thm: intervening id} demonstrates that P-OR, P-IPW, and P-hybrid formulas developed for the PIIE in Section~\ref{sec: proximal PIIE identification} are robust enough to identify the effect of an intervening variable $A_M$, even when interventions on exposure $A$ are ill-defined. Critically, unlike the proximal inference for the NIE \citep{dukes2023proximal} or PIIE, this approach does not require untestable cross-world independence conditions (e.g., Assumption~\ref{asm: latent cross world}). The proximal intervening variable framework offers a more empirically grounded basis for causal mediation inference.

\section{Semiparametric Inference}

The identification results in Theorems~\ref{thm: proximal mediation}-\ref{thm: proximal mediation-h&q} and \ref{thm: intervening id} establish a foundation for estimation using proxy variables to overcome the challenges posed by pervasive unmeasured confounding and ill-defined exposure interventions. However, standard plug-in estimators often suffer from first-order bias due to the estimation of infinite-dimensional nuisance functions. To achieve robustness, we derive the efficient influence function (EIF) of the target estimands. This facilitates the construction of estimators with second-order bias properties, a critical strategy for valid inference in complex semiparametric models \citep{robins2008higher, robins2017Minimax}. 

To unify our analysis, let $\Psi$ denote the \textit{oracle generalized front-door functional}:
\begin{equation*}
    \Psi := \iiiint \E\left [ Y|U=u,A=a,M=m,X=x \right ] \mathrm{d}F(m|u,A=0,x) \mathrm{d}F(a|u,x) \mathrm{d}F(u|x)\mathrm{d}F(x).
\end{equation*}
Under Assumptions~\ref{asm: positivity}-\ref{asm: latent cross world}, the PIIE functional $\psi$ simplifies to $\Psi$ as established in \eqref{eq: PIIE-front door oracle}. Similarly, under Assumptions~\ref{asm: positivity} and \ref{asm: intervening consistency}-\ref{asm: latent dismissible}, the intervening variable estimand $\E\left[ Y(a_M = 0) \right]$ is also equivalent to $\Psi$, as indicated in \eqref{eq: intervening-front door oracle}. Consequently, the following semiparametric theories apply broadly to both the proximal mediation and intervening variable frameworks. Our proposed estimators are asymptotically normal, multiply robust to nuisance function misspecification, and achieve local efficiency when all nuisance functions are correctly specified.

\vspace{-0.2cm}
\subsection{Semiparametric efficiency bound}

In this section, we analyze the inference for $\Psi$ under the semiparametric model $\mathcal{M}_{sp}$, which imposes no restrictions on the observed data distribution beyond the existence (but not necessarily uniqueness) of outcome confounding bridge functions $h_1$ and $h_0$ solving \eqref{eq: h1} and \eqref{eq: h0}. Since the existence of these bridge functions restricts the tangent space, the semiparametric efficiency bound can be derived under specific regularity conditions. Let $T_1: L_2(W,M,A,X) \rightarrow L_2(Z,A,M,X)$ and $T_0: L_2(W,M,X) \rightarrow L_2(Z,A=0,X)$ be conditional expectation operators defined by $T_1(g) = \E\left [ g(W,M,A,X)|Z,A,M,X \right ]$ and $T_0(g) = \E\left [ g(W,M,X)|Z,A=0,X \right ]$.

\begin{assumption}\emph{(Regularity)}:
    $T_1$ and $T_0$ are surjective at the true data-generating mechanism.
\label{asm: T1 surjective}
\end{assumption}

As noted by \cite{ying2023proximal}, Assumption~\ref{asm: T1 surjective} implies that the functions in the domain are sufficiently rich to generate any element in the target space via $T_1$ and $T_0$. Under these conditions, we establish the following efficiency bound for estimating $\Psi$, which is distinct from the bound for the NIE considered by \cite{dukes2023proximal}.

\begin{theorem}
    Suppose that confounding bridge functions $h_1$ and $h_0$ exist for all data-generating laws in $\mathcal{M}_{sp}$. Further, assume that at the true data-generating law, there exist confounding bridge functions $q_0$ and $q_1$ solving \eqref{eq: q0} and \eqref{eq: q1}, and that Assumption~\ref{asm: completeness} holds. Then, the following is a valid influence function for $\Psi$ under $\mathcal{M}_{sp}$:
    \begin{align*}
        EIF_{\Psi} =& I(A=0) q_0(Z,X) \left [ h_1(W,M,1,X)-h_0(W,1,X) \right ]\\& + I(A=0) \left [ h_1(W,M,0,X)-h_0(W,0,X) \right ] \\
        &+ \left [ I(A=1) q_1(Z, M, X) + I(A=0) \right ] \left [ Y-h_1(W,M,A,X) \right ] + h_0(W,A,X) - \Psi.
    \end{align*}
    Moreover, $EIF_{\Psi}$ is the EIF in the submodel under Assumption~\ref{asm: T1 surjective} and the uniqueness of all bridge functions; the corresponding efficiency bound is $\E(EIF_{\Psi}^2)$.
\label{thm: efficiency bound}
\end{theorem}

\subsection{Multiply robust and locally efficient estimation}\label{sec: robust}

The estimator for $\Psi$ derived from the EIF in Theorem~\ref{thm: efficiency bound} exhibits multiple robustness. It remains consistent if one of the following three submodels of $\cM_{sp}$ is correctly specified:
\begin{align*}
    &\pmb{\cM_1:} h_1(W,M,A,X) ~\text{and}~ h_0(W,A,X) ~\text{are correctly specified}; \\
    &\pmb{\cM_2:} h_1(W,M,A,X) ~\text{and}~ q_0(Z,X) ~\text{are correctly specified}; \\
    &\pmb{\cM_3:} q_1(Z,M,X) ~\text{and}~ q_0(Z,X) ~\text{are correctly specified}.
\end{align*}
Consistency is guaranteed under the union model $\cM_{union} = \cM_1 \cup \cM_2 \cup \cM_3$, without requiring the researcher to identify which submodel holds.

Suppose the bridge functions $h_1$, $h_0$, $q_0$, $q_1$ are parameterized by finite-dimensional vectors $\beta_1$, $\beta_0$, $\gamma_0$, and $\gamma_1$, respectively. We obtain nuisance estimates $(\hat{\beta}_1, \hat{\beta}_0)$ and $\left( \hat{\gamma}_0, \hat{\gamma}_1 \right)$ by solving empirical estimating equations derived from the integral definitions \eqref{eq: h1}-\eqref{eq: h0} and \eqref{eq: q0_integral}-\eqref{eq: q1_integral}:
\begin{align}
    \sum_{i=1}^{n} \left [ Y_i - h_1(W_i,M_i,A_i,X_i; \beta_1) \right ] c_1(Z_i,M_i,A_i,X_i) =& 0, \label{eq: h1_solve} \\
    \sum_{i=1}^{n} (1-A_i) \left [ h_1(W_i,M_i,a,X_i; \beta_1) - h_0(W_i,a,X_i; \beta_0) \right ] c_{0a}(Z_i,X_i) =& 0, ~\text{for}~ a\in\left \{ 0,1 \right \}, \label{eq: h0_solve} \\
    \sum_{i=1}^{n} \left [ (1-A_i) q_0(Z_i,X_i; \gamma_0) - A_i \right ] d_0(W_i,X_i) =& 0, \label{eq: q0_solve} \\
    \sum_{i=1}^{n} \left [ A_i q_1(Z_i,M_i,X_i;\gamma_1) - (1-A_i)q_0(Z_i,X_i;\hat{\gamma}_0) \right ] d_1(W_i, M_i, X_i) =& 0, \label{eq: q1_solve}
\end{align}
Here, the functions $c_1(Z,M,A,X)$, $d_0(W,X)$, and $d_1(W,M,X)$ share the same dimensions as their corresponding parameters $\beta_1$, $\gamma_0$, and $\gamma_1$. By partitioning $\beta_0 = (\beta_{01}, \beta_{00})$, we can express $h_0(W,A,X; \beta_0) = A h_{0}(W,1,X; \beta_{01}) + (1-A) h_{0}(W,0,X; \beta_{00})$, with $c_{0a}(Z,X)$ taking the same dimension as $\beta_{0a}$ for $a \in \left \{ 0,1 \right \}$. While Equations~\eqref{eq: q0} and \eqref{eq: q1} might initially suggest that estimating $\gamma_0$ and $\gamma_1$ requires modeling $f(A=1|W, X)/f(A=0|W, X)$ and $f(A=0|W,M,X)/f(A=1|W,M,X)$, Proposition~\ref{propo_integral} demonstrates this is unnecessary. Efficient choices for the functions $c_1, c_{0a}$, $d_0$, and $d_1$ are provided in the Appendix of \cite{cui2023semiparametric}. However, because the efficiency gain over simple linear bases, such as $c_1(Z, M, A, X)=(1, Z, M, A, X)^T, c_{0a}(Z, X)=(1, Z, X)^T$, $d_0(W, X)=(1, W, X)^T$, and $d_1(W, M, X) = (1, W, M, X)^T$, is typically modest in most scenarios \citep{stephens2014locally}, we do not pursue locally efficient estimation of these nuisance parameters further.

We can then construct the proximal outcome regression (P-OR), proximal hybrid (P-hybrid), and proximal inverse probability weighting (P-IPW) estimators of $\Psi$ as follows:
\begin{align*}
    \hat{\Psi}_{P-OR} =& \frac{1}{n} \sum_{i=1}^{n} h_0(W_i,A_i,X_i; \hat{\beta}_0), \\
    \hat{\Psi}_{P-hybrid} =& \frac{1}{n} \sum_{i=1}^{n} \left [ (1-A_i)q_0(Z_i,X_i; \hat{\gamma}_0) h_1(W_i,M_i,1,X_i; \hat{\beta}_1) + (1-A_i)h_1(W_i,M_i,0,X_i; \hat{\beta}_1) \right ], \\
    \hat{\Psi}_{P-IPW} =& \frac{1}{n} \sum_{i=1}^{n} \left [ A_i q_1(Z_i,M_i,X_i; \hat{\gamma}_1) Y_i + (1-A_i) Y_i \right ].
\end{align*}
Specifically, $\hat{\Psi}_{P-OR}$ is consistent and asymptotically normal (CAN) under submodel $\cM_1$, $\hat{\Psi}_{P-hybrid}$ is CAN under submodel $\cM_2$, and $\hat{\Psi}_{P-IPW}$ is CAN under submodel $\cM_3$. To unify these properties, we develop a proximal multiply robust (P-MR) estimator for $\Psi$, which relaxes parametric assumptions and attains the semiparametric efficiency bound locally.

\begin{theorem}\emph{(\textbf{P-MR}):}
  The estimator $\hat{\Psi}_{P-MR}$ defined as:
   \begin{align*}
       \hat{\Psi}_{P-MR} =& \frac{1}{n} \sum_{i=1}^{n} \left\{ (1-A_i) q_0(Z_i,X_i; \hat{\gamma}_0) \left [ h_1(W_i,M_i,1,X_i; \hat{\beta}_1)-h_0(W_i,1,X_i; \hat{\beta}_0) \right ] \right. \\
       &+ (1-A_i) \left [ h_1(W_i,M_i,0,X_i; \hat{\beta}_1)-h_0(W_i,0,X_i; \hat{\beta}_0) \right ] \\
       &+ \left. \left [ A_i q_1(Z_i,M_i,X_i; \hat{\gamma}_1) + 1-A_i \right ] \left [ Y_i - h_1(W_i,M_i,A_i,X_i; \hat{\beta}_1) \right ] + h_0(W_i,A_i,X_i; \hat{\beta}_0) \right\}
   \end{align*}
   is a consistent and asymptotically normal estimator of $\Psi$ under the union model $\cM_{union} = \cM_1 \cup \cM_2 \cup \cM_3$. Furthermore, it is semiparametric locally efficient in $\cM_{sp}$ at the intersection submodel $\cM_{int} = \cM_1 \cap \cM_2 \cap \cM_3$, provided Assumption~\ref{asm: T1 surjective} holds.
\label{thm: multiply robust}
\end{theorem}

A key advantage of this EIF-based estimator is its Neyman orthogonality \citep{neyman1979c}, making it first-order insensitive to nuisance parameter estimation errors. This second-order bias property accommodates the slower $o\left ( n^{-1/4} \right )$ convergence rates typical of modern machine learning for the underlying nuisance parameters in high-dimensional settings \citep{chernozhukov2018double, kennedy2024semiparametric, li2025identification}. As detailed in Theorem~\ref{thm: debiased ML} in the Appendix~\ref{sec: debiased ML}, integrating $EIF_{\Psi}$ with a debiased machine learning cross-fitting framework \citep{schick1986asymptotically, chernozhukov2018double} preserves $\sqrt{n}$-consistency and asymptotic normality. This guarantees valid inference even when nuisance bridge functions are estimated via modern machine learning approaches, such as the adversarial regularized minimax learning \citep{dikkala2020minimax, kallus2021causal, ghassami2022minimax}.

\section{Numerical Experiments}\label{sec: simulation}

We evaluate the finite-sample performance of the proposed estimators under model misspecification; the full data-generating process is detailed in the Appendix~\ref{app: DGP}. The target estimand is $\Psi$, representing either the PIIE functional $\psi$ (under well-defined interventions on $A$) or the intervening variable estimand $\E\left[ Y(a_M = 0) \right]$. 

We compare our four proximal estimators ($\hat{\Psi}_{P-OR}$, $\hat{\Psi}_{P-hybrid}$, $\hat{\Psi}_{P-IPW}$, $\hat{\Psi}_{P-MR}$) against the non-proximal doubly robust (DR) estimator \citep{fulcher2020robust} that assumes no unmeasured confounding affects the mediator $M$. Confounding bridge functions are estimated via Equations~\eqref{eq: h1_solve}-\eqref{eq: q1_solve} under the nuisance models that are specified and justified in the Appendix~\ref{app: numerical experiments}. We assess performance across four scenarios: 
\begin{itemize}
    \item \textbf{Scenario 1:} All models correctly modeled ($\cM_1$, $\cM_2$, $\cM_3$ hold).
    \item \textbf{Scenario 2:} Exposure bridge functions $q_1$, $q_0$ misspecified (only $\cM_1$ holds).
    \item \textbf{Scenario 3:} Bridge functions $q_1$, $h_0$ misspecified (only $\cM_2$ holds).
    \item \textbf{Scenario 4:} Outcome bridge functions $h_1$, $h_0$ misspecified (only $\cM_3$ holds).
\end{itemize}
Model misspecification is induced by replacing the covariate vector $X = \left(X_1, X_2 \right)^T$ with the non-linear transformation $X^{\ast} = \left( \sqrt{\left | X_1 \right |} +3, \sqrt{\left | X_2 \right |} +3 \right)^T$ during estimation. Results are based on sample size $n = 1000$ and $500$ Monte Carlo iterations, with confidence intervals constructed via nonparametric bootstrap.

\begin{table}[p]
	\centering
    \resizebox{0.7\textwidth}{!}{
		\begin{tabular}{cccccccccc}
			\toprule
			Scenarios & \multicolumn{3}{c}{Models} &  & \multicolumn{5}{c}{Estimators} \\
			\cmidrule{2-4}
			\cmidrule{6-10}
			&\textbf{$\mathcal{M}_1$} & \textbf{$\mathcal{M}_2$} & \textbf{$\mathcal{M}_3$} & & \textit{DR} & \textit{P-OR} & \textit{P-hybrid} & \textit{P-IPW} & \textit{P-MR} \\
			\cmidrule{2-4}
			\cmidrule{6-10}
			\multicolumn{10}{c}{ \textbf{Bias} } \\
			1 & \checkmark & \checkmark & \checkmark & & $-0.09$ & $\mathbf{-0.00}$ & $\mathbf{-0.00}$ & $\mathbf{-0.00}$ & $\mathbf{-0.01}$ \\
			2 & \checkmark & $\times$ & $\times$ & & $-0.09$ & $\mathbf{0.00}$ & $0.23$ & $0.12$ & $\mathbf{0.01}$ \\
			3 & $\times$ & \checkmark & $\times$ & & $-0.09$ & $0.09$ & $\mathbf{-0.00}$ & $-0.05$ & $\mathbf{0.00}$ \\
			4 & $\times$ & $\times$ & \checkmark & & $-0.10$ &$0.11$ & $0.07$ & $\mathbf{0.00}$ & $\mathbf{-0.00}$ \\
			\\
			\multicolumn{10}{c}{ \textbf{MSE} } \\
			1 & \checkmark & \checkmark & \checkmark & & $0.01$ & $\pmb{0.01}$ & $\pmb{0.01}$ & $\pmb{0.01}$ & $\pmb{0.01}$ \\
			2 & \checkmark & $\times$ & $\times$ & & $0.01$ & \pmb{$0.01$} & $0.06$ & $0.03$ & \pmb{$0.01$} \\
			3 & $\times$ & \checkmark & $\times$ & & $0.01$ & $0.02$ & \pmb{$0.01$} & $0.11$ & $\pmb{0.02}$ \\
			4 & $\times$ & $\times$ & \checkmark & & $0.01$ & $0.02$ & $0.02$ & \pmb{$0.01$} & $\pmb{0.01}$ \\
            \\
			\multicolumn{10}{c}{ \textbf{Coverage} } \\
			1 & \checkmark & \checkmark & \checkmark & & $33.2\%$ & \pmb{$98.2\%$} & \pmb{$98.0\%$} & \pmb{$95.4\%$} & \pmb{$96.4\%$} \\
			2 & \checkmark & $\times$ & $\times$ & & $33.6\%$ & \pmb{$98.4\%$} & $9.8\%$ & $46.6\%$ & \pmb{$95.6\%$} \\
			3 & $\times$ & \checkmark & $\times$ & & $30.4\%$ & $50.8\%$ & \pmb{$97.4\%$} & $41.4\%$ & \pmb{$95.0\%$} \\
			4 & $\times$ & $\times$ & \checkmark & & $28.2\%$ & $42.4\%$ & $72.2\%$ & \pmb{$95.6\%$} & \pmb{$95.2\%$} \\
            \\
			\multicolumn{10}{c}{ \textbf{Length} } \\
			1 & \checkmark & \checkmark & \checkmark & & $0.15$ & \pmb{$0.37$} & \pmb{$0.40$} & \pmb{$0.38$} & \pmb{$0.42$} \\
			2 & \checkmark & $\times$ & $\times$ & & $0.15$ & \pmb{$0.38$} & $0.42$ & $0.39$ & \pmb{$0.44$} \\
			3 & $\times$ & \checkmark & $\times$ & & $0.15$ & $0.37$ & \pmb{$0.40$} & $0.39$ & \pmb{$0.43$} \\
			4 & $\times$ & $\times$ & \checkmark & & $0.15$ & $0.38$ & $0.41$ & \pmb{$0.40$} & \pmb{$0.44$} \\
			\bottomrule
			\end{tabular}}
		\caption{Simulation results across Scenarios 1-4. Bias: Monte Carlo bias; MSE: mean squared error; Coverage: $95\%$ confidence interval coverage rate; Length: average $95\%$ confidence interval length. Emboldened values denote consistent estimators in the corresponding scenarios.}
    \label{table: Experiments 1-4}
\end{table}

Table~\ref{table: Experiments 1-4} summarizes the results. Due to the unmeasured confounder $U$, the non-proximal DR estimator exhibits substantial bias and severe undercoverage (approximately $31\%$). As expected, the basic proximal estimators perform well only when their corresponding submodels are correct: P-OR succeeds in Scenarios 1 and 2, P-hybrid in Scenarios 1 and 3, and P-IPW in Scenarios 1 and 4. In contrast, the proximal multiply robust estimator P-MR empirically validates semiparametric theory, maintaining negligible bias and nominal coverage (over $95\%$) across all four scenarios. This highlights the P-MR estimator's robust utility for maintaining valid inference despite complex, varying model misspecifications.

\section{Data Application}

In Example~\ref{example: alcohol consumption}, evaluating interventions that encourage hazardous alcohol consumption is ethically questionable; thus, we focus on the PIIE rather than the NIE. We apply our proximal estimators to survey data from $n=3856$ Mexican Health Ministry workers collected during the COVID-19 pandemic \cite{dominguez2023mediation}. We investigate the indirect effect of alcohol consumption on depression, mediated by depersonalization symptoms. Alcohol consumption is measured using the WHO Alcohol Use Disorders Identification Test (AUDIT) \citep{babor2001audit}, while depression is assessed via the CESD-47 scale \citep{franco2018prevalence}. The burnout scale \citep{maslach1981measurement} provides a 12-item assessment across three dimensions: emotional exhaustion, achievement dissatisfaction, and depersonalization.

In our analysis, the exposure $A=1$ denotes hazardous alcohol consumption, indicated by scoring $\geq 1$ on AUDIT questions 2 or 3. The mediator $M$ is a composite measure for depersonalization-related symptoms following the burnout scale, and the outcome $Y$ is a composite measure for depression-related symptoms following the CESD-47 scale. Standard approaches, such as the generalized front-door formula \citep{fulcher2020robust, wen2024causal}, fail in this setting because they strictly require the absence of unmeasured confounding affecting the mediator. This is biologically and psychologically implausible: unmeasured factors like genetic predisposition or early-life trauma ($U$) may directly affect $A$, $M$, and $Y$ simultaneously, while environmental stressors ($V$) may confound the $A-Y$ relationship.

To account for $U$, we use baseline covariates as proxy variables. Alcohol dependence and peer influence on drinking affect $M$ and $Y$ strictly through $A$; thus, we assign them as exposure-inducing confounding proxies ($Z$). Emotional exhaustion and achievement dissatisfaction do not directly affect $A$ \citep{dominguez2023mediation}, and they, as well as depersonalization, represent the three dimensions of burnout \citep{maslach1981measurement}; thus, they are associated with $A$, $M$, and $Z$ only through unmeasured confounders, and we categorize them as outcome-inducing confounding proxies ($W$). Standard errors and $95\%$ confidence intervals (CIs) for all estimators are calculated using the nonparametric bootstrap.

\begin{table}[ht]
    \centering
    \begin{tabular}{ccccc}
         \hline
         & P-OR & P-hybrid & P-IPW & P-MR \\
         \hline
         Exposure indirect effects (SEs) & $0.12~\left(0.08\right)$ & $0.17~\left(0.13\right)$ & $-0.08~\left(0.08\right)$ & $\pmb{0.20}~\left(0.14\right)$ \\
         \hline
         95\% CIs & $\left ( 0.05, 0.38 \right )$ & $\left ( -0.11, 0.40 \right )$ & $\left ( -0.27, 0.03 \right )$ & $\pmb{\left ( 0.06, 0.61 \right )}$ \\
         \hline
         \\
    \end{tabular}
    \caption{Exposure effect estimates (standard errors) and $95\%$ confidence intervals for the PIIE of alcohol consumption on depression, mediated by depersonalization.}
    \label{table: real data}
\end{table}

Table~\ref{table: real data} summarizes the results. The P-OR, P-hybrid, and P-MR estimators indicate a positive population intervention indirect effect, whereas the P-IPW yields a negative estimate. Notably, only the P-OR and P-MR estimates are statistically significant (their 95\% CIs exclude zero). The concordance between the P-OR and P-MR approaches strengthens confidence that the bridge functions $h_1$ and $h_0$ (semiparametric submodel $\cM_1$) are correctly specified. Ultimately, the P-MR results suggest that an increased level of alcohol consumption indirectly elevates the risk of depression via depersonalization symptoms.

\section{Comparison with \cite{dukes2023proximal} and \cite{ghassami2024causal}}

This paper investigates the mediating role of $M$ within the causal pathway from an exposure $A$ and an outcome $Y$. While the natural indirect effect (NIE) and the population intervention indirect effect (PIIE) both quantify causal mediation, the PIIE offers distinct advantages for policy evaluations involving exposures that cannot be ethically manipulated. Critically, the PIIE framework facilitates a formal interpretation of the causal effect of an intervening variable under ill-defined exposure interventions, a robust feature that the NIE typically lacks. We construct proximal causal inference frameworks that allow for unmeasured confounding across the $A-Y$, $A-M$, and $M-Y$ relationships. Under such pervasive confounding, standard identification strategies, including Pearl's front-door criterion \citep{pearl2009causality} and the generalized front-door formula \citep{fulcher2020robust, wen2024causal}, are generally inapplicable. Recent studies \citep{dukes2023proximal, ghassami2024causal} have also developed methods that leverage proxy variables to mitigate confounding bias in mediation analysis.

\tikzstyle{var} = [circle, very thick,draw=black,fill=white,minimum size=10mm]
\tikzstyle{var_unob} = [circle, very thick,draw=black,fill=gray,minimum size=10mm]
\tikzstyle{var_uninter} = [circle, dashed,very thick,draw=black,fill=white,minimum size=10mm]
\begin{figure}[ht]
    \centering
       \subfloat[\label{fig comparison: Dukes}]{
        \resizebox{0.23\textwidth}{!}{
       \begin{tikzpicture}[>=latex, scale=0.7, every node/.style={scale=0.7}]
       \node (A) [var, xshift=-3cm,yshift=0cm]{$\boldsymbol{A}$};
       \node (M) [var, xshift=0cm,yshift=-2cm] {$\boldsymbol{M}$};
       \node (Y) [var, xshift=3cm,yshift=0cm]{$\boldsymbol{Y}$};
       \node (X) [var, xshift=0cm,yshift=2cm]{$\boldsymbol{X}$};
       \node (Z) [var, xshift=-2.5cm,yshift=3cm]
       {$\boldsymbol{Z}$};
       \node (W) [var, xshift=2.5cm,yshift=3cm]
       {$\boldsymbol{W}$};
       \node (U) [var_unob, xshift=0cm,yshift=4cm]{$\boldsymbol{U}$};
       \draw[->, very thick] (X) -- (A);
       \draw[->, very thick] (A) -- (Y);
       \draw[->, very thick] (X) -- (M);
       \draw[->, very thick] (X) -- (Z);
       \draw[->, very thick] (X) -- (W);
       \draw[->, very thick] (Z) -- (A);
       \draw[->, very thick] (W) -- (Y);
       \draw[->, very thick] (A) -- (M);
       \draw[->, very thick] (M) -- (Y);
       \draw[->, very thick] (X) -- (Y);
       \draw[->, very thick] (U) -- (Y);
       \draw[->, very thick] (U) -- (A);
       \draw[->, very thick] (U) -- (X);
       \draw[->, very thick] (U) -- (X);
       \draw[->, very thick] (U) -- (Z);
       \draw[->, very thick] (U) -- (W);
       \draw[->, very thick] (-0.4,3.7) arc (100:255: 1 and 3);
       \end{tikzpicture}
         }
       }
       \subfloat[\label{fig comparison: Ghassami}]{
        \resizebox{0.23\textwidth}{!}{
        \begin{tikzpicture}
        [>=latex, scale=0.7, every node/.style={scale=0.7}]
       \node (A) [var, xshift=-3cm,yshift=1cm]{$\boldsymbol{A}$};
       \node (M) [var_unob, xshift=0cm,yshift=-2cm] {$\boldsymbol{M}$};
       \node (Y) [var, xshift=3cm,yshift=1cm]{$\boldsymbol{Y}$};
       \node (X) [var, xshift=0cm,yshift=2cm]{$\boldsymbol{X}$};
       \node (Z) [var, xshift=-2.5cm,yshift=-1.5cm]
       {$\boldsymbol{Z}$};
       \node (W) [var, xshift=2.5cm,yshift=-1.5cm]
       {$\boldsymbol{W}$};
       \node (U) [var_unob, xshift=0cm,yshift=4cm]{$\boldsymbol{U}$};
       \draw[->, very thick] (X) -- (A);
       \draw[->, very thick] (A) -- (Y);
       \draw[->, very thick] (X) -- (M);
       \draw[->, very thick] (X) -- (Z);
       \draw[->, very thick] (X) -- (W);
       \draw[->, very thick] (A) -- (Z);
       \draw[->, very thick] (W) -- (Y);
       \draw[->, very thick] (A) -- (M);
       \draw[->, very thick] (M) -- (Y);
       \draw[->, very thick] (Z) -- (M);
       \draw[->, very thick] (M) -- (W);
       \draw[->, very thick] (X) -- (Y);
       \draw[->, very thick] (U) -- (Y);
       \draw[->, very thick] (U) -- (A);
       \draw[->, very thick] (U) -- (X);
       \draw[->, very thick] (U) -- (X);
       \end{tikzpicture}
         }
       }  
       \subfloat[\label{fig comparison: PIIE}]{
        \resizebox{0.23\textwidth}{!}{
         \begin{tikzpicture}[>=latex, scale=0.7, every node/.style={scale=0.7}]
       \node (A) [var, xshift=-3cm,yshift=0cm]{$\boldsymbol{A}$};
       \node (M) [var, xshift=0cm,yshift=-2cm] {$\boldsymbol{M}$};
       \node (Y) [var, xshift=3cm,yshift=0cm]{$\boldsymbol{Y}$};
       \node (X) [var, xshift=0cm,yshift=2cm]{$\boldsymbol{X}$};
       \node (Z) [var, xshift=-2.5cm,yshift=3cm]
       {$\boldsymbol{Z}$};
       \node (W) [var, xshift=2.5cm,yshift=3cm]
       {$\boldsymbol{W}$};
       \node (U) [var_unob, xshift=0cm,yshift=4cm]{$\boldsymbol{U}$};
       \node (V) [var_unob, xshift=0cm,yshift=6cm]{$\boldsymbol{V}$};
       \draw[->, very thick] (X) -- (A);
       \draw[->, very thick] (A) -- (Y);
       \draw[->, very thick] (X) -- (M);
       \draw[->, very thick] (X) -- (Z);
       \draw[->, very thick] (X) -- (W);
       \draw[->, very thick] (Z) -- (A);
       \draw[->, very thick] (W) -- (Y);
       \draw[->, very thick] (A) -- (M);
       \draw[->, very thick] (M) -- (Y);
       \draw[->, very thick] (X) -- (Y);
       \draw[->, very thick] (U) -- (Y);
       \draw[->, very thick] (U) -- (A);
       \draw[->, very thick] (U) -- (X);
       \draw[->, very thick] (U) -- (X);
       \draw[->, very thick] (U) -- (Z);
       \draw[->, very thick] (U) -- (W);
       \draw[->, very thick] (V) -- (U);
       \draw[->, very thick] (V) -- (A);
       \draw[->, very thick] (V) -- (Y);
       \draw[->, very thick] (-0.4,3.7) arc (100:255: 1 and 3);
       \end{tikzpicture}
        }
     }
     \subfloat[\label{fig comparison: intervening variable}]{
      \resizebox{0.23\textwidth}{!}{
       \begin{tikzpicture}[>=latex, scale=0.7, every node/.style={scale=0.7}]
       \node (A) [var_uninter, xshift=-3cm,yshift=0cm]{$\boldsymbol{A}$};
       \node (A_M) [var,  xshift=-2cm,yshift=-1.8cm]{$\boldsymbol{A_M}$};
       \node (M) [var, xshift=0cm,yshift=-2cm] {$\boldsymbol{M}$};
       \node (Y) [var, xshift=3cm,yshift=0cm]{$\boldsymbol{Y}$};
       \node (X) [var, xshift=0cm,yshift=2cm]{$\boldsymbol{X}$};
       \node (Z) [var, xshift=-2.5cm,yshift=3cm]
       {$\boldsymbol{Z}$};
       \node (W) [var, xshift=2.5cm,yshift=3cm]
       {$\boldsymbol{W}$};
       \node (U) [var_unob, xshift=0cm,yshift=4cm]{$\boldsymbol{U}$};
       \node (V) [var_unob, xshift=0cm,yshift=6cm]{$\boldsymbol{V}$};
       \draw[->, line width = 2.8pt] (A) -- (A_M);
       \draw[->, very thick] (A_M) -- (M);
       \draw[->, very thick] (X) -- (A);
       \draw[->, very thick] (A) -- (Y);
       \draw[->, very thick] (X) -- (M);
       \draw[->, very thick] (X) -- (Z);
       \draw[->, very thick] (X) -- (W);
       \draw[->, very thick] (Z) -- (A);
       \draw[->, very thick] (W) -- (Y);
       \draw[->, very thick] (M) -- (Y);
       \draw[->, very thick] (X) -- (Y);
       \draw[->, very thick] (U) -- (Y);
       \draw[->, very thick] (U) -- (A);
       \draw[->, very thick] (U) -- (X);
       \draw[->, very thick] (U) -- (X);
       \draw[->, very thick] (U) -- (Z);
       \draw[->, very thick] (U) -- (W);
       \draw[->, very thick] (V) -- (U);
       \draw[->, very thick] (V) -- (A);
       \draw[->, very thick] (V) -- (Y);
       \draw[<-, very thick] (0.3,-1.6) arc (285:430 : 1 and 2.8);
       \end{tikzpicture}
       }
    }
    \caption{(a) a DAG representing the structural assumptions for proximal mediation as established in \cite{dukes2023proximal}; (b) a DAG consistent with the identification framework in \cite{ghassami2024causal}; (c) a DAG illustrating our proposed model assumptions; (d) an extended DAG incorporating ill-defined exposure interventions. In all panels, gray vertices indicate unobserved variables, the bold arrow signifies a deterministic relationship, and the interventions on the dashed vertex are ill-defined.}
    \label{fig: comparison}
\end{figure}

\cite{dukes2023proximal} consider the decomposition of the average causal effect (ACE) as follows:
\begin{align*}
    \E\left [ Y(1) - Y(0) \right ] =& \E\left [ Y(1,M(1)) - Y(0,M(0)) \right ] \\
    =& \underbrace{ \E\left [ Y(1,M(1)) - Y(1,M(0)) \right ] }_{\text{Natural Indirect Effect (NIE)}} + \underbrace{ \E\left [ Y(1,M(0)) - Y(0,M(0)) \right ] }_{\text{Natural Direct Effect (NDE)}},
\end{align*}
with a primary focus on the estimation of the NIE. Their identification strategy relies on proxy variables that capture the entirety of the confounding mechanism; specifically, they assume that proxies $Z$ and $W$ account for the unmeasured confounder $U$ under an extra latent conditional exchangeability assumption, $Y(a,m) \indep A|U, X$ (see Figure~\ref{fig comparison: Dukes}). In contrast, we focus on the decomposition of the population intervention effect (PIE) and develop a proximal causal inference framework for the PIIE. Notably, our approach relaxes these requirements by only necessitating proxies for a subset of the confounding mechanism. Specifically, our model accommodates an additional unmeasured confounder $V$ affecting the $A-Y$ relationship that is not necessarily captured by $Z$ and $W$ (see Figures~\ref{fig comparison: PIIE} and \ref{fig comparison: intervening variable}).

While \cite{ghassami2024causal} also targets the PIIE, they consider a specialized setting where the mediator $M$ is latent and assume this unobserved mediator is directly unaffected by unmeasured confounders. In their framework, proxies are leveraged for the hidden mediator itself rather than the unmeasured confounding mechanism (see Figure~\ref{fig comparison: Ghassami}). Conversely, our approach addresses the more typical mediation setting with an observed mediator $M$. We further weaken the assumptions on the unmeasured confounding mechanism by allowing for pervasive unmeasured confounding across the $A-M$, $M-Y$, and $A-Y$ relationships (see Figures~\ref{fig comparison: PIIE} and \ref{fig comparison: intervening variable}).

Figure~\ref{fig comparison: PIIE} illustrates our proposed proximal framework for PIIE inference. A key advantage of our approach is that it maintains causal interpretability even when interventions on the exposure $A$ are ill-defined (see Figure~\ref{fig comparison: intervening variable}). In such contexts, the estimand can be formally interpreted as the causal effect of an intervening variable $A_M$. Our models accommodate unmeasured common causes $U$ of $\left \{ A,M,Y \right \}$ and $V$ of $\left \{ A,Y \right \}$, requiring proxies $Z$ and $W$ only for $U$. Furthermore, the framework for the intervening variable effect (Figure~\ref{fig comparison: intervening variable}) sidesteps the cross-world independence assumptions. These assumptions are empirically untestable but remain a necessary requirement for standard mediation models \citep[e.g.,][]{dukes2023proximal, ghassami2024causal} targeting the NIE or PIIE.

\section*{Acknowledgement}

Yifan Cui and Baoluo Sun are co-corresponding authors. Yifan Cui was partially supported by the National Key R\&D Program of China (2024YFA1015600), the National Natural Science Foundation of China (12471266 and U23A2064). Baoluo Sun was partially supported by Singapore MOE AcRF Tier 1 grants (A-8000452-00-00, A-8002935-00-00).

\newpage
\appendix
\begin{center}
\LARGE Appendix
\end{center}

\section{Completeness conditions}\label{app: completeness}

Completeness is a technical condition linked to sufficiency in statistical inference and is critical for identification in nonparametric regression with instrumental variables \citep{newey2003instrumental}. In our setting, Assumption~\ref{asm: completeness} connects the range of $U$ to that of $Z$, ensuring that any variation in $U$ is fully captured by the corresponding variation in $Z$ when $A=0$ and $X$ are fixed. This requires $Z$ to exhibit sufficient variability relative to $U$, given $A, M$, and $X$. For the case of categorical $U$ and $Z$ with number of categories $\textbf{\#}_u$ and $\textbf{\#}_z$, respectively, Assumption~\ref{asm: completeness} implies $\textbf{\#}_z \geq \textbf{\#}_u$,
meaning that proxies must have at least as many categories as the unmeasured confounders. For continuously distributed confounders, completeness holds under many common distributions. Suppose that the distribution of $U$ given $Z$, $A=0$, and $X$ is absolutely continuous, with density
\begin{equation}
    f(u|Z=z, A=0, X=x) = s(u) \exp\left\{\eta(z, x)^T t(u) - \alpha(z, x) \right\},
    \label{eq: exponential family}
\end{equation}
where $s(u)>0$, $t(u)$ is a one-to-one function of $u$, and the support of $\eta(z, x)$ is an open set; then the completeness assumption~\ref{asm: completeness} holds \citep{newey2003instrumental}. Notably, model~\eqref{eq: exponential family} represents the exponential family with parameters $z$ and $x$. The justifications for completeness have also been discussed for location-scale families \citep{hu2018nonparametric}, as well as nonparametric instrumental regression models \citep{d2011completeness, darolles2011nonparametric}. Thus, measuring a comprehensive set of proxy variables can be a strategy to control for unmeasured confounding.

\section{Details about the intervening variable causal model \citep{wen2024causal}}\label{app: intervening variable}

In the mediation model of \cite{wen2024causal}, we consider the challenge that interventions on exposure $A$ are ill-defined. To investigate the indirect effect of $A$ on $Y$ through $M$ in this challenge, we turn to the analysis of a modifiable intervening variable $A_M$, such that (i) $A_M$ is deterministically equal to $A$ in the observed data and (ii) $A_M$ fully captures the effect of $A$ on $Y$ through $M$. Similar conditions have also been considered for literature on separable effects \citep{robins2010alternative, robins2022interventionist, stensrud2021generalized, stensrud2023conditional}. The counterfactuals induced by interventions on $A_M$ can also be defined analogously; for example, let $M(a_M)$ and $Y(a_M)$ denote the counterfactual mediator and potential outcome, respectively, had the intervening variable been set to $A_M = a_M$.

Ill-defined interventions are major factors in the replication crisis for policy-relevant estimands in medicine and public health. When an analysis targets exposures that do not correspond to well-defined interventions, future experiments cannot operationalize these variables in the same way as the initial analysis, rendering replication impossible. For example, it is prevalent to explore the indirect causal effect of chronic pain $A$ on mortality $Y$ through opioid prescription $M$ \citep{inoue2022causal}. However, it is unclear how one could intervene in chronic pain, or even whether any intervention on chronic pain can be well-defined in policy-making. Hence, the estimand regarding interventions on chronic pain has dubious public health implications \citep{holland1986statistics}. In contrast, \cite{wen2024causal} suggest exploring the causal effect of an intervening variable, which corresponds to the part of interventions that are feasible to implement in practice. Specifically, consider a doctor who determines if a patient should receive opioids to relieve chronic pain, whose perception of the chronic pain status is deterministically the same as whether the patient suffers from chronic pain. Note that the doctor could be asked to consider a patient as (or not) having chronic pain in their perception, even if it is opposite to the real chronic pain status; interventions on the doctor's perception are well-defined. In this example, the doctor's perception of the patient's chronic pain status is a modifiable intervening variable $A_M$, which is entirely determined by chronic pain and fully captures the effect of chronic pain on opioid prescription.

For the binary exposure case, the intervening variable estimand is defined as the expected counterfactual outcome $\E\left[ Y(a_M) \right]$, for $a_M \in \left \{ 0, 1 \right \}$. As such, a causal effect of an intervening variable $A_M$ is defined as a contrast between $\E\left[ Y(a_M=0) \right]$ and another causal estimand such as $\E\left[ Y(a_M=1) \right]$ or $\E(Y)$. In the example of chronic pain and opioid prescription considered by \cite{inoue2022causal} and \cite{wen2024causal}, note that chronic pain status should no longer be used to determine opioid prescription according to the current guidelines by the Centers for Disease Control and Prevention (CDC) \citep{dowell2016cdc,inoue2022causal}, then there is interest in evaluating this modified prescription policy, by comparing the natural perception on opioid prescription with the suggested policy that the doctor disregards a patient's chronic pain in their decisions on opioid prescription. Hence, $\E\left[ Y - Y(a_M=0) \right]$ should be of primary interest in this example. 

With a slight abuse of notation from \cite{stensrud2021generalized}, let the superscript `G' refer to a future trial where $A_M$ is randomly assigned. To identify the intervening variable estimand, \cite{wen2024causal} impose the following conditions:
$\text{(N1)}:\text{If}~ A_M=a_M, \text{then}~ M(a_M) = M;
~\text{If}~ A_M=a_M, \text{then}~ Y(a_M) = Y;
\text{(N2)}: A = A_M;
\text{(N3)}: Y^{G} \indep A_M^{G} | A^{G}, M^{G}, L^{G}; M^{G} \indep A^{G} | A_M^{G}, L^{G}.$ 

N1 states that interventions on the intervening variable $A_M$ are well-defined, which accommodates interventions on the exposure $A$ to be ill-defined. N2 implies that $A_M$ is deterministically equal to $A$. N3 is the so-called dismissible component condition assumed in the scenario of `G', substantially ensuring the absence of unmeasured confounding in both $M-Y$ and $A-M$ relations, while allowing for an unmeasured common cause $V$ of $A$ and $Y$. Additionally, N3 would fail if there is a direct effect of $A_M$ on $Y$ not mediated by $M$, or $A$ exerts an effect on $M$ not through $A_M$. Under conditions N1-N3, the intervening variable estimand $\E\left[ Y(a_M) \right]$ is identified by the generalized front-door formula \citep{wen2024causal}, that is,
\begin{equation*}
    \E\left[ Y(a_M) \right] = \iiint \E\left [ Y|A=a,M=m,L=l \right ] \mathrm{d}F(m|A=a_M,l) \mathrm{d}F(a|l) \mathrm{d}F(l).
\end{equation*}

Note that although both $\psi$ and $\E\left[ Y(a_M) \right]$ can be identified by the generalized front-door formula, their identification conditions are clearly distinct. On the one hand, the consistency condition M1 for $\psi$ would be violated in the example of opioid prescription, where it is unclear how to specify a well-defined intervention on chronic pain. On the other hand, the cross-world exchangeability condition like M3 is untestable in principle \citep{dawid2000causal, robins2010alternative} but is widely required to identify mediation estimands, even for the natural indirect effect \citep{Pearl2001direct}, whereas it is unnecessary for the identification of intervening variable estimands. In contrast, the analysis of intervening variable estimands relies on the existence of a well-defined $A_M$ satisfying N1-N3, equipped with good interpretability and realistic research meaning.

\section{Proofs of identification results}

In this section, we provide proofs of the main identification results.

\subsection{Proof of Proposition~\ref{propo_integral}}

Starting from \eqref{eq: q0}, we obtain that
\begin{align*}
    &\frac{f(A=1|W,X)}{f(A=0|W,X)} = \E\left [ q_0(Z,X)|W,A=0,X \right ] \\
    \Longleftrightarrow \quad& \E\left [ q_0(Z,X)|W,A=0,X \right ] f(A=0|W,X) = f(A=1|W,X) \\
    \Longleftrightarrow \quad& \E\left \{ (1-A) \E\left [ q_0(Z,X)|W,A=0,X \right ] - A \mid W,X \right \} = 0 \\
    \Longleftrightarrow \quad& \E\left [ (1-A) q_0(Z,X) - A |W,X \right ] = 0.
\end{align*}

Similarly, it follows from \eqref{eq: q1} that
{\small \begin{align*}
    &\E\left[ q_0(Z,X)|W,A=0,M,X \right] \frac{f(A=0|W,M,X)}{f(A=1|W,M,X)} = \E\left[ q_1(Z,M,X)|W,A=1,M,X \right] \\
    \Longleftrightarrow \quad& \E\left[ q_0(Z,X)|W,A=0,M,X \right] f(A=0|W,M,X) = \E\left[ q_1(Z,M,X)|W,A=1,M,X \right] f(A=1|W,M,X) \\
    \Longleftrightarrow \quad& \E\left \{ (1-A) \E\left[ q_0(Z,X)|W,A=0,M,X \right] \mid W,M,X \right \} = \E\left \{ A \E\left[ q_1(Z,M,X)|W,A=1,M,X \right] \mid W,M,X \right \} \\
    \Longleftrightarrow \quad&  \E\left[ (1-A) q_0(Z,X)|W,M,X \right] = \E\left[ A q_1(Z,M,X)|W,M,X \right].
\end{align*}}

\subsection{Proof of Theorem~\ref{thm: proximal mediation}}\label{app: proof of PIIE proximal}

Before illustrating the main proof of identification, we expound that the integral equations~\eqref{eq: h1}-\eqref{eq: h0} imply \eqref{eq: h1-U}-\eqref{eq: h0-U}:
\begin{align}
    \E(Y|U,A,M,X) =& \int h_1(w,M,A,X) \mathrm{d}F(w|U,A,M,X), \label{eq: h1-U} \\
    \E\left [ h_1(W,M,a,X)|U,A=0,X \right ] =& \int h_0(w,a,X) \mathrm{d}F(w|U,A=0,X), \label{eq: h0-U}
\end{align}

We start with the integral equation~\eqref{eq: h1} to prove Equation~\eqref{eq: h1-U}. The left hand side of Equation~\eqref{eq: h1} equals to
\begin{align*}
    \E(Y|Z,A,M,X) =& \E\left [  \E(Y|U,Z,A,M,X) |Z,A,M,X \right ] \\
    =& \E\left [  \E(Y|U,A,M,X) |Z,A,M,X \right ]. \qquad (Y \indep Z|U, A, M, X)
\end{align*}
Also, the right hand side of Equation~\eqref{eq: h1} is equal to
\begin{align*}
    \int h_1(w,M,A,X) \mathrm{d}F(w|Z,A,M,X) =& \iint h_1(w,M,A,X) \mathrm{d}F(w|u,Z,A,M,X) \mathrm{d}F(u|Z,A,M,X) \\
    =& \iint h_1(w,M,A,X) \mathrm{d}F(w|u,A,M,X) \mathrm{d}F(u|Z,A,M,X),
\end{align*}
where the second equality sign is due to $W \indep Z|U, A, M, X$. Furthermore, by the completeness assumption~\ref{asm: completeness}, we obtain Equation~\eqref{eq: h1-U} that
\begin{equation*}
    \E(Y|U,A,M,X) = \int h_1(w,M,A,X) \mathrm{d}F(w|U,A,M,X).
\end{equation*}

Then, we give the proof of Equation~\eqref{eq: h0-U}. The left hand side of Equation~\eqref{eq: h0} equals to 
\begin{align*}
     \E\left [ h_1(W,M,a,X)|Z,A=0,X \right ] =& \E\left \{  \E\left [ h_1(W,M,a,X)|U,Z,A=0,X \right ] |Z,A=0,X \right \} \\
     =& \E\left \{ \E\left [ h_1(W,M,a,X)|U,A=0,X \right ] |Z,A=0,X \right \}.  ~\left(Z \indep (W,M)|U, A, X \right)
\end{align*}
Also, the right hand side of Equation~\eqref{eq: h0} is equal to
\begin{align*}
    \int h_0(w,a,X) \mathrm{d}F(w|Z,A=0,X) =& \iint h_0(w,a,X) \mathrm{d}F(w|u,Z,A=0,X) \mathrm{d}F(u|Z,A=0,X) \\
    =& \iint h_0(w,a,X) \mathrm{d}F(w|u,A=0,X) \mathrm{d}F(u|Z,A=0,X),
\end{align*}
where the second equality sign is due to $W \indep Z|U, A, X$. Next, by the completeness assumption~\ref{asm: completeness}, we obtain Equation~\eqref{eq: h0-U} that
\begin{equation*}
    \E\left [ h_1(W,M,a,X)|U,A=0,X \right ] = \int h_0(w,a,X) \mathrm{d}F(w|U,A=0,X).
\end{equation*}

Finally, it follows from the above results that
{\small \begin{align*}
    \E\left[Y(A, M(0))\right] =& \int \E\left[Y(A, M(0)) |u,A=a,x \right] \mathrm{d}F(u,a,x) \\
    =& \iint \E\left[Y(A, M(0)) |u,A=a,M(0)=m,x \right] \mathrm{d}F_{M(0)}(m|u,a,x) \mathrm{d}F(u,a,x) \\
    =& \iint \E\left[Y(a,m) |u,A=a,M(0)=m,x \right] \mathrm{d}F_{M(0)}(m|u,a,x) \mathrm{d}F(u,a,x) \\
    =& \iint \E\left[Y(a,m) |u,A=a,x \right] \mathrm{d}F_{M(0)}(m|u,a,x) \mathrm{d}F(u,a,x) \qquad \text{(Assumption~\ref{asm: latent cross world})} \\
    =& \iint \E\left[Y(a,m) |u,A=a,M=m,x \right] \mathrm{d}F_{M(0)}(m|u,a,x) \mathrm{d}F(u,a,x) \qquad \text{(Assumption~\ref{asm: latent cross world})} \\
    =& \iint \E\left[Y |u,A=a,M=m,x \right] \mathrm{d}F_{M(0)}(m|u,a,x) \mathrm{d}F(u,a,x) \qquad \text{(Assumption~\ref{asm: consistency}\textcolor{blue}{.2})} \\
    =& \iint \E\left[Y |u,A=a,M=m,x \right] \mathrm{d}F_{M(0)}(m|u,A=0,x) \mathrm{d}F(u,a,x) \qquad \text{(Assumption~\ref{asm: latent exchangeability})} \\
    =& \iint \E\left[Y |u,A=a,M=m,x \right] \mathrm{d}F(m|u,A=0,x) \mathrm{d}F(u,a,x) \qquad \text{(Assumption~\ref{asm: consistency}\textcolor{blue}{.1})} \\
    =& \iiint h_1(w,m,a,x) \mathrm{d}F(w|u,a,m,x) \mathrm{d}F(m|u,A=0,x) \mathrm{d}F(u,a,x) \qquad \text{(Equation~\eqref{eq: h1-U})} \\
    =& \iiint h_1(w,m,a,x) \mathrm{d}F(w|u,A=0,m,x) \mathrm{d}F(m|u,A=0,x) \mathrm{d}F(u,a,x) ~(W \indep A|U, M, X) \\
    =& \iint h_1(w,m,a,x) \mathrm{d}F(w,m|u,A=0,x) \mathrm{d}F(u,a,x) \\
    =& \iint h_0(w,a,X) \mathrm{d}F(w|u,A=0,x) \mathrm{d}F(u,a,x) \qquad \text{(Equation~\eqref{eq: h0-U})} \\
    =& \iint h_0(w,a,X) \mathrm{d}F(w|u,a,x) \mathrm{d}F(u,a,x) \qquad (W \indep A|U, X) \\
    =& \iint h_0(w,a,X) \mathrm{d}F(w,a|x) \mathrm{d}F(x),
\end{align*}}
where we index subscript by $M(0)$ to indicate that the distribution refers to the counterfactual $M(0)$ rather than the observed $M$.

\subsection{Proofs of Theorems~\ref{thm: proximal mediation-q} and \ref{thm: proximal mediation-h&q}}\label{app: proof of PIIE proximal-q}

Firstly, we focus on the proof of Theorem~\ref{thm: proximal mediation-h&q}. We start from the integral equation~\eqref{eq: q0} to prove Equation~\eqref{eq: q0-U}:
\begin{equation}
    \frac{f(A=1|U,X)}{f(A=0|U,X)} = \E\left [ q_0(Z,X)|U,A=0,X \right ]. \label{eq: q0-U}
\end{equation}
Note that
\begin{align*}
     \frac{f(A=1|W,X)}{f(A=0|W,X)} =& \frac{1}{f(A=0|W,X)} - 1 \\
     =& \frac{f(W,X)}{f(A=0,W,X)} - 1 \\
     =& \int \frac{f(u,W,X)}{f(A=0,W,X)} \mathrm{d}u - 1 \\
     =& \int \frac{f(u|A=0,W,X)}{f(A=0|u,W,X)} \mathrm{d}u - 1 \\
     =& \int \frac{1}{f(A=0|u,W,X)} \mathrm{d}F(u|A=0,W,X) - 1 \\
     =& \int \frac{f(A=1|u,W,X)}{f(A=0|u,W,X)} \mathrm{d}F(u|A=0,W,X) \\
     =& \int \frac{f(A=1|u,X)}{f(A=0|u,X)} \mathrm{d}F(u|A=0,W,X), \qquad (W \indep A|U,X)
\end{align*}
as well as
\begin{align*}
    \frac{f(A=1|W,X)}{f(A=0|W,X)} =& \E\left [ q_0(Z,X)|W,A=0,X \right ] \\
    =& \int \E\left [ q_0(Z,X)|u,W,A=0,X \right ] \mathrm{d}F(u|W,A=0,X) \\
    =& \int \E\left [ q_0(Z,X)|u,A=0,X \right ] \mathrm{d}F(u|W,A=0,X), \qquad (Z \indep W|U,A,X)
\end{align*}
it follows from Assumption~\ref{asm: completeness-q}\textcolor{blue}{.2} that
\begin{equation*}
    \frac{f(A=1|U,X)}{f(A=0|U,X)} = \E\left [ q_0(Z,X)|U,A=0,X \right ].
\end{equation*}
Next, the proof of Theorem~\ref{thm: proximal mediation} implies Equation~\eqref{eq: h1-U} under the relevant conditions. Furthermore,
{\small \begin{align*}
    \E\left[Y(A, M(0))\right] =& \iiint \E\left[Y |u,A=a,M=m,x \right] \mathrm{d}F(m|u,A=0,x) \mathrm{d}F(a|u,x) \mathrm{d}F(u,x) \\
    =& \iint \E\left[Y |u,A=1,M=m,x \right] f(A=1|u,x) \mathrm{d}F(m|u,A=0,x) \mathrm{d}F(u,x) \\
    &+ \iint \E\left[Y |u,A=0,M=m,x \right] f(A=0|u,x) \mathrm{d}F(m|u,A=0,x) \mathrm{d}F(u,x) \\
    =& \E\left \{ \E\left [ \E(Y |U,A=1,M,X)|U,A=0,X \right ] f(A=1|U,X) \right \} \\
    &+ \E\left \{ \E\left [ \E(Y |U,A=0,M,X)|U,A=0,X \right ] f(A=0|U,X) \right \} \\
    =& \E\left \{ \E\left [ \frac{I(A=0)}{f(A=0|U,X)}\E(Y |U,A=1,M,X)|U,X \right ] f(A=1|U,X) \right \} \\
    &+ \E\left \{ \E\left [ \frac{I(A=0)}{f(A=0|U,X)} \E(Y |U,A=0,M,X)|U,X \right ] f(A=0|U,X) \right \} \\
    =& \E\left [ I(A=0) \frac{f(A=1|U,X)}{f(A=0|U,X)}\E(Y |U,A=1,M,X) \right ] + \E\left [ I(A=0) \E(Y |U,A=0,M,X) \right ]
\end{align*}}
where we refer to the proof of Theorem~\ref{thm: proximal mediation} to obtain the first equality under Assumptions~\ref{asm: consistency}, \ref{asm: latent exchangeability}, and \ref{asm: latent cross world}. The first term is equal to
\begin{align*}
    &\E\left [ I(A=0) \frac{f(A=1|U,X)}{f(A=0|U,X)}\E(Y |U,A=1,M,X) \right ] \\
    =& \E\left [ I(A=0) \E(q_0(Z,X)|U,A=0,X) \E(Y |U,A=1,M,X) \right ] \qquad (\text{Equation}~\eqref{eq: q0-U}) \\
    =& \E\left [ I(A=0) \E(q_0(Z,X)|U,A=0,M,X) \E(Y |U,A=1,M,X) \right ] \qquad (Z \indep M |U,A,X) \\
    =& \E\left [ I(A=0) q_0(Z,X) \E(Y |U,A=1,M,X) \right ] \\
    =& \E\left [ I(A=0) q_0(Z,X) \E(h_1(W,M,1,X)|U,A=1,M,X) \right ] \qquad (\text{Equation}~\eqref{eq: h1-U}) \\
    =& \E\left [ I(A=0) q_0(Z,X) \E(h_1(W,M,1,X)|U,A,Z,M,X) \right ] \qquad (W \indep (Z,A)|U, M, X) \\
    =& \E\left [ I(A=0) q_0(Z,X) h_1(W,M,1,X) \right ].
\end{align*}
The second term turns out to be
\begin{align*}
    &\E\left [ I(A=0) \E(Y |U,A=0,M,X) \right ] \\
    =& \E\left [ I(A=0) \E(h_1(W,M,0,X)|U,A=0,M,X) \right ]  \qquad (\text{Equation}~\eqref{eq: h1-U}) \\
    =& \E\left [ I(A=0) h_1(W,M,0,X) \right ].
\end{align*}
Finally, combining the two terms gives the identification equation~\eqref{eq: PIIE id-q0}:
\begin{equation*}
    \E\left[Y(A, M(0))\right] = \E\left [ I(A=0) q_0(Z,X) h_1(W,M,1,X) + I(A=0) h_1(W,M,0,X) \right ].
\end{equation*}

For the proof of Theorem~\ref{thm: proximal mediation-q}, we start from the integral equation~\eqref{eq: q1} to prove Equation~\eqref{eq: q1-U}:
\begin{equation}
    \E\left[ q_0(Z,X)|U,A=0,M,X \right] \frac{f(A=0|U,M,X)}{f(A=1|U,M,X)} = \E\left[ q_1(Z,M,X)|U,A=1,M,X \right]. \label{eq: q1-U}
\end{equation}
First,
\begin{align*}
    &\E\left[ q_0(Z,X)|W,A=0,M,X \right] \frac{f(A=0|W,M,X)}{f(A=1|W,M,X)} \\
    =& \frac{1}{f(A=1|W,M,X)} \int \E\left[ q_0(Z,X)|u,W,A=0,M,X \right] f(A=0|W,M,X) \mathrm{d}F(u|W,A=0,M,X) \\
    =& \frac{1}{f(A=1|W,M,X)} \int \E\left[ q_0(Z,X)|u,W,A=0,M,X \right] f(A=0|u,W,M,X) \mathrm{d}F(u|W,M,X) \\
    =& \int \E\left[ q_0(Z,X)|u,W,A=0,M,X \right] \frac{f(A=0|u,W,M,X)}{f(A=1|u,W,M,X)} \mathrm{d}F(u|W,A=1,M,X) \\
    =& \int \E\left[ q_0(Z,X)|u,W,A=0,M,X \right] \frac{f(A=0|u,M,X)}{f(A=1|u,M,X)} \mathrm{d}F(u|W,A=1,M,X) \qquad (W \indep A|U,M,X) \\
    =& \int \E\left[ q_0(Z,X)|u,A=0,M,X \right] \frac{f(A=0|u,M,X)}{f(A=1|u,M,X)} \mathrm{d}F(u|W,A=1,M,X), \qquad (Z \indep W|U,A,M,X) 
\end{align*}
where for the second and third equal signs, we employ $f(A=0|W, M, X) f(u|W, A=0, M, X) = f(A=0|u, W, M, X) f(u|W, M, X)$ and $\frac{f(u|W, M, X)}{f(A=1|W, M, X)} = \frac{f(u|W, A=1, M, X)}{f(A=1|u, W, M, X)}$, respectively. In addition,
\begin{align*}
    &\E\left[ q_0(Z,X)|W,A=0,M,X \right] \frac{f(A=0|W,M,X)}{f(A=1|W,M,X)} \\
    =& \E\left[ q_1(Z,M,X)|W,A=1,M,X \right] \\
    =& \int \E\left[ q_1(Z,M,X)|u,W,A=1,M,X \right] \mathrm{d}F(u|W,A=1,M,X) \\
    =& \int \E\left[ q_1(Z,M,X)|u,A=1,M,X \right] \mathrm{d}F(u|W,A=1,M,X). \qquad (Z \indep W|U,A,M,X)
\end{align*}
Hence, it follows from Assumption~\ref{asm: completeness-q}\textcolor{blue}{.1} that
\begin{equation*}
    \E\left[ q_0(Z,X)|U,A=0,M,X \right] \frac{f(A=0|U,M,X)}{f(A=1|U,M,X)} = \E\left[ q_1(Z,M,X)|U,A=1,M,X \right].
\end{equation*}
Finally, the identification equation~\eqref{eq: PIIE id-q1} is indicated by
{\small \begin{align*}
    &\E\left[Y(A, M(0))\right] \\
    =& \E\left [ I(A=0) \frac{f(A=1|U,X)}{f(A=0|U,X)}\E(Y |U,A=1,M,X) \right ] + \E\left [ I(A=0) \E(Y |U,A=0,M,X) \right ] \\
    =& \E\left [ I(A=0) \E(q_0(Z,X)|U,A=0,M,X) \E(Y |U,A=1,M,X) \right ] + \E\left [ I(A=0) \E(Y |U,A=0,M,X) \right ] \\
    =& \E\left [ \E(q_0(Z,X)|U,A=0,M,X) \E(Y |U,A=1,M,X) f(A=0|U,M,X) \right ] + \E\left [ I(A=0) \E(Y |U,A,M,X) \right ] \\
    =& \E\left [ \E(q_0(Z,X)|U,A=0,M,X) \E\left(\frac{I(A=1)}{f(A=1|U,M,X)}Y |U,M,X \right) f(A=0|U,M,X) \right ] + \E\left [ I(A=0) Y \right ] \\
    =& \E\left [ I(A=1) \frac{f(A=0|U,M,X)}{f(A=1|U,M,X)}Y \E(q_0(Z,X)|U,A=0,M,X) \right ] + \E\left [ I(A=0) Y \right ] \\
    =& \E\left [ I(A=1) Y \E(q_1(Z,M,X)|U,A=1,M,X) \right ] + \E\left [ I(A=0) Y \right ] \\
    =& \E\left [ I(A=1) Y \E(q_1(Z,M,X)|Y,U,A,M,X) \right ] + \E\left [ I(A=0) Y \right ] \\
    =& \E\left [ I(A=1) Y q_1(Z,M,X) + I(A=0) Y \right ],
\end{align*}}
where we refer to the first part of this proof to obtain the first equality, the second equality is due to Equation~\eqref{eq: q0-U} and $Z \indep M |U, A, X$, the sixth equality is implied by Equation~\eqref{eq: q1-U}, and the seventh equality results from $Z \indep Y |U, A, M, X$.

\subsection{Proof of Theorem~\ref{thm: intervening id}}

To simplify notation, we use $\E_G$ and $F_G$ to denote the expectations and distributions in the G scenario. Under Assumptions~\ref{asm: positivity} and \ref{asm: intervening consistency}-\ref{asm: latent dismissible}, we obtain that
{\small \begin{align}
     \E\left[ Y(a_M = 0) \right] =& \E_G\left[ Y(a_M = 0) \right] \notag \\
     =& \int \E_G\left[ Y(a_M = 0)|U=u,A=a,X=x \right] dF_G(u,a,x) \notag \\
     =& \int \E_G\left[ Y(a_M = 0)|U=u,A=a,A_{M}=0,X=x \right] dF_G(u,a,x) \notag \\
     =& \int \E_G\left[ Y|U=u,A=a,A_{M}=0,X=x \right] dF_G(u,a,x) \notag \\
     =& \iint \E_G\left[ Y|U=u,A=a,A_{M}=0,M=m,X=x \right] dF_G(m|u,A=a,A_M=0,x) dF_G(u,a,x) \notag \\
     =& \iint \E_G\left[ Y|U=u,A=a,A_{M}=a,M=m,X=x \right] dF_G(m|u,A=0,A_M=0,x) dF_G(u,a,x) \notag \\
     =& \iint \E\left[ Y|U=u,A=a,A_{M}=a,M=m,X=x \right] dF(m|u,A=0,A_M=0,x) dF(u,a,x) \notag \\
     =&  \iint \E\left[ Y|U=u,A=a,M=m,X=x \right] dF(m|u,A=0,x) dF(u,a,x), \label{eq: intervening oracle front-door}
\end{align}}
where the third equality follows from the definition of the G scenario in which $A_M$ is randomly assigned, the fourth equality is due to the intervening consistency Assumption~\ref{asm: intervening consistency}, the sixth equality is implied by the latent dismissible component Assumption~\ref{asm: latent dismissible} such that $Y^{G} \indep A_M^{G} |U^{G}, A^{G}, M^{G}, X^{G}$ and $M^{G} \indep A^{G} |U^{G}, A_M^{G}, X^{G}$, and the last equality results from the intervening variable determinism Assumption~\ref{asm: intervening determinism} such that the event $\left \{ A=a, A_M=a \right \}$ is equivalent to $\left \{ A=a \right \}$.

Note that the proofs in Appendix~\ref{app: proof of PIIE proximal} imply that under Assumptions~\ref{asm: proxies}-\ref{asm: completeness}, the confounding bridge functions $h_1(W, M, A, X)$ and $h_0(W, A, X)$ that satisfy Assumption~\ref{asm: outcome bridge functions} are also solutions of Equations~\eqref{eq: h1-U}-\eqref{eq: h0-U}. Based on the above results, we obtain the proof of identification~\eqref{eq: intervening id}: 
{\small\begin{align*}
    \E\left[ Y(a_M = 0) \right] =& \iint \E\left[ Y|U=u,A=a,M=m,X=x \right] dF(m|u,A=0,x) dF(u,a,x) \qquad (\text{Equation}~\eqref{eq: intervening oracle front-door}) \\
    =& \iiint h_1(w,m,a,x) \mathrm{d}F(w|u,a,m,x) \mathrm{d}F(m|u,A=0,x) \mathrm{d}F(u,a,x) \qquad (\text{Equation}~\eqref{eq: h1-U}) \\
    =& \iiint h_1(w,m,a,x) \mathrm{d}F(w|u,A=0,m,x) \mathrm{d}F(m|u,A=0,x) \mathrm{d}F(u,a,x) \qquad (W \indep A|U, M, X) \\
    =& \iint h_1(w,m,a,x) \mathrm{d}F(w,m|u,A=0,x) \mathrm{d}F(u,a,x) \\
    =& \iint h_0(w,a,X) \mathrm{d}F(w|u,A=0,x) \mathrm{d}F(u,a,x) \qquad (\text{Equation}~\eqref{eq: h0-U}) \\
    =& \iint h_0(w,a,X) \mathrm{d}F(w|u,a,x) \mathrm{d}F(u,a,x) \qquad (W \indep A|U, X) \\
    =& \iint h_0(w,a,X) \mathrm{d}F(w,a|x) \mathrm{d}F(x)
\end{align*}}

Next, we turn to the proof of identification~\eqref{eq: intervening id-q0}. Note that the proofs in Appendix~\ref{app: proof of PIIE proximal-q} imply that under Assumptions~\ref{asm: proxies}, part (i) of \ref{asm: completeness}, and part (i) of \ref{asm: completeness-q}, the confounding bridge functions $h_1(W, M, A, X)$ satisfying part (i) of Assumption~\ref{asm: outcome bridge functions} and $q_0(Z, X)$ satisfying part (i) of Assumption~\ref{asm: exposure bridge functions} are solutions of Equations~\eqref{eq: h1-U} and \eqref{eq: q0-U}, respectively. Hence, 
{\small \begin{align*}
    &\E\left[ Y(a_M = 0) \right] \\ 
    =& \iint \E\left[ Y|U=u,A=a,M=m,X=x \right] dF(m|u,A=0,x) dF(u,a,x) \qquad (\text{Equation}~\eqref{eq: intervening oracle front-door}) \\
    =& \iint \E\left[Y |u,A=1,M=m,x \right] f(A=1|u,x) \mathrm{d}F(m|u,A=0,x) \mathrm{d}F(u,x) \\
    &+ \iint \E\left[Y |u,A=0,M=m,x \right] f(A=0|u,x) \mathrm{d}F(m|u,A=0,x) \mathrm{d}F(u,x) \\
    =& \E\left \{ \E\left [ \E(Y |U,A=1,M,X)|U,A=0,X \right ] f(A=1|U,X) \right \} \\
    &+ \E\left \{ \E\left [ \E(Y |U,A=0,M,X)|U,A=0,X \right ] f(A=0|U,X) \right \} \\
    =& \E\left \{ \E\left [ \frac{I(A=0)}{f(A=0|U,X)}\E(Y |U,A=1,M,X)|U,X \right ] f(A=1|U,X) \right \} \\
    &+ \E\left \{ \E\left [ \frac{I(A=0)}{f(A=0|U,X)} \E(Y |U,A=0,M,X)|U,X \right ] f(A=0|U,X) \right \} \\
    =& \E\left [ I(A=0) \frac{f(A=1|U,X)}{f(A=0|U,X)}\E(Y |U,A=1,M,X) \right ] + \E\left [ I(A=0) \E(Y |U,A=0,M,X) \right ] \\
    =& \E\left [ I(A=0) \E(q_0(Z,X)|U,A=0,X) \E(Y |U,A=1,M,X) \right ] + \E\left [ I(A=0) \E(Y |U,A=0,M,X) \right ] \qquad (\text{Equation}~\eqref{eq: q0-U}) \\
    =& \E\left [ I(A=0) \E(q_0(Z,X)|U,A=0,M,X) \E(Y |U,A=1,M,X) \right ] + \E\left [ I(A=0) \E(Y |U,A=0,M,X) \right ] ~(Z \indep M |U,A,X) \\
    =& \E\left [ I(A=0) q_0(Z,X) \E(Y |U,A=1,M,X) \right ] + \E\left [ I(A=0) \E(Y |U,A=0,M,X) \right ] \\
    =& \E\left [ I(A=0) q_0(Z,X) \E(h_1(W,M,1,X)|U,A=1,M,X) \right ] + \E\left [ I(A=0) \E(h_1(W,M,0,X)|U,A=0,M,X) \right ] \qquad \eqref{eq: h1-U} \\
    =& \E\left [ I(A=0) q_0(Z,X) \E(h_1(W,M,1,X)|U,A,Z,M,X) \right ] + \E\left [ I(A=0) h_1(W,M,0,X) \right ] \qquad (W \indep (Z,A)|U, M, X) \\
    =& \E\left [ I(A=0) q_0(Z,X) h_1(W,M,1,X) + I(A=0) h_1(W,M,0,X) \right ].
\end{align*}}

Finally, we focus on the proof of identification~\eqref{eq: intervening id-q1}. Note that the proofs in Appendix~\ref{app: proof of PIIE proximal-q} imply that under Assumptions~\ref{asm: proxies} and \ref{asm: completeness-q}, the confounding bridge functions $q_0(Z, X)$ and $q_1(Z, M, X)$ that satisfy Assumption~\ref{asm: exposure bridge functions} are also solutions of Equations~\eqref{eq: q0-U}-\eqref{eq: q1-U}. Therefore, 
{\small \begin{align*}
    &\E\left[ Y(a_M = 0) \right] \\ 
    =& \iint \E\left[ Y|U=u,A=a,M=m,X=x \right] dF(m|u,A=0,x) dF(u,a,x) \qquad (\text{Equation}~\eqref{eq: intervening oracle front-door}) \\
    =& \E\left [ I(A=0) \frac{f(A=1|U,X)}{f(A=0|U,X)}\E(Y |U,A=1,M,X) \right ] + \E\left [ I(A=0) \E(Y |U,A=0,M,X) \right ] \\
    =& \E\left [ I(A=0) \E(q_0(Z,X)|U,A=0,M,X) \E(Y |U,A=1,M,X) \right ] + \E\left [ I(A=0) \E(Y |U,A=0,M,X) \right ] \\
    =& \E\left [ \E(q_0(Z,X)|U,A=0,M,X) \E(Y |U,A=1,M,X) f(A=0|U,M,X) \right ] + \E\left [ I(A=0) \E(Y |U,A,M,X) \right ] \\
    =& \E\left [ \E(q_0(Z,X)|U,A=0,M,X) \E\left(\frac{I(A=1)}{f(A=1|U,M,X)}Y |U,M,X \right) f(A=0|U,M,X) \right ] + \E\left [ I(A=0) Y \right ] \\
    =& \E\left [ I(A=1) \frac{f(A=0|U,M,X)}{f(A=1|U,M,X)}Y \E(q_0(Z,X)|U,A=0,M,X) \right ] + \E\left [ I(A=0) Y \right ] \\
    =& \E\left [ I(A=1) Y \E(q_1(Z,M,X)|U,A=1,M,X) \right ] + \E\left [ I(A=0) Y \right ] \\
    =& \E\left [ I(A=1) Y \E(q_1(Z,M,X)|Y,U,A,M,X) \right ] + \E\left [ I(A=0) Y \right ] \\
    =& \E\left [ I(A=1) Y q_1(Z,M,X) + I(A=0) Y \right ],
\end{align*}}
where we refer to the proof of \eqref{eq: intervening id-q0} to obtain the second equality, the third equality is due to Equation~\eqref{eq: q0-U} and $Z \indep M |U, A, X$, the seventh equality is implied by Equation~\eqref{eq: q1-U}, and the eighth equality follows from $Z \indep Y |U, A, M, X$.

\section{Proofs of semiparametric theories}

\subsection{Proof of Theorem~\ref{thm: efficiency bound}}\label{app: proof of efficiency bound}

In order to first obtain an influence function for $\Psi$ under the semiparametric model $\mathcal{M}_{sp}$, one needs to find the mean zero random variable $G$ for which
\begin{equation}
    \frac{\partial \Psi_t}{\partial t}|_{t=0} = \E\left [ G S(\mathcal{O};t) \right ]|_{t=0},
\label{eq: pathwise derivative}
\end{equation}
where $\mathcal{O}$ represents the observed full data $\mathcal{O} = \left ( Y, W, Z, A, M, X \right )$, $S(\mathcal{O};t) = \partial \log f(\mathcal{O};t)/\partial t$ is the score, and $\Psi_t$ is the parameter of interest $\Psi$ under a one-dimensional regular parametric submodel in $\mathcal{M}_{sp}$ indexed by $t$ that included the true data generating mechanism at $t=0$ \citep{van2000asymptotic}.

Recall the moment restrictions
\begin{align*}
    \E\left[Y - h_1(W,M,A,X)|Z,A,M,X\right] =& 0, \\
    \E\left [ h_1(W,M,a,X) - h_0(W,a,X)|Z,A=0,X \right ] =& 0,
\end{align*}
for $a=0, 1$, implied by integral equations~\eqref{eq: h1} and \eqref{eq: h0}. Hence we take derivatives with respect to regular parametric submodels indexed by $t$ in $\mathcal{M}_{sp}$ and get
\begin{align*}
    \partial\E_{t}\left[Y - h_{1t}(W,M,A,X)|Z,A,M,X\right]/\partial t|_{t=0} =& 0, \\
     \partial\E_{t}\left [ h_{1t}(W,M,a,X) - h_{0t}(W,a,X)|Z,A=0,X \right ]/\partial t|_{t=0} =& 0.
\end{align*}
Let $\varepsilon_1 = Y - h_1(W,M,A,X)$ and $\varepsilon_{0a} = h_1(W,M,a,X) - h_0(W,a,X)$, for $a=0, 1$, the above equations turn out to be
\begin{align}
    \E\left[ \frac{\partial}{\partial t} h_{1t}(W,M,A,X)\big|_{t=0}|Z,A,M,X\right] = \E\left [ \varepsilon_1 S(Y,W|Z,A,M,X) |Z,A,M,X \right ], \label{eq: pathwise h1}
\end{align}
and
\begin{align}
    &\E\left[ \frac{\partial}{\partial t} h_{0t}(W,a,X)\big|_{t=0}|Z,A=0,X\right] \nonumber \\
    =& \E\left[ \frac{\partial}{\partial t} h_{1t}(W,M,a,X)\big|_{t=0}|Z,A=0,X\right] + \E\left [ \varepsilon_{0a} S(W,M|Z,A=0,X) |Z,A=0,X \right ]. \label{eq: pathwise h0}
\end{align}

Furthermore, the left hand side of Equation~\eqref{eq: pathwise derivative} in $\mathcal{M}_{sp}$ is equal to
\begin{align*}
    \frac{\partial \Psi_t}{\partial t}|_{t=0} =& \frac{\partial}{\partial t} \E_t\left [ h_{0t}(W,A,X) \right ]\big|_{t=0} \\
    =& \E\left [ h_{0}(W,A,X) S(W,A,X) \right ] + \E\left [ \frac{\partial}{\partial t} h_{0t}(W,A,X)\big|_{t=0} \right ] \\
    =& \E\left \{ [h_{0}(W,A,X) - \Psi] S(\mathcal{O}) \right \} + \E\left [ \frac{\partial}{\partial t} h_{0t}(W,A,X)\big|_{t=0} \right ].
\end{align*}
For the second term, we have that
{\small \begin{align*}
    &\E\left [ \frac{\partial}{\partial t} h_{0t}(W,A,X)\big|_{t=0} \right ] = \E\left \{ \E\left [ \frac{\partial}{\partial t} h_{0t}(W,A,X)\big|_{t=0} |W,X \right ] \right \} \\
    =& \E\left \{ \frac{\partial}{\partial t} h_{0t}(W,1,X)\big|_{t=0} f(A=1|W,X) + \frac{\partial}{\partial t} h_{0t}(W,0,X)\big|_{t=0} f(A=0|W,X) \right \} \\
    =& \E\left \{ \frac{I(A=0)}{f(A=0|W,X)}  f(A=1|W,X) \frac{\partial}{\partial t} h_{0t}(W,1,X)\big|_{t=0} + I(A=0) \frac{\partial}{\partial t} h_{0t}(W,0,X)\big|_{t=0} \right \} \\
    =& \E\left \{ I(A=0) \E\left [ q_0(Z,X)|W,A=0,X \right ] \frac{\partial}{\partial t} h_{0t}(W,1,X)\big|_{t=0} + I(A=0) \frac{\partial}{\partial t} h_{0t}(W,0,X)\big|_{t=0} \right \} \quad (\text{Equation}~\eqref{eq: q0}) \\
    =& \E\left \{ I(A=0) q_0(Z,X) \frac{\partial}{\partial t} h_{0t}(W,1,X)\big|_{t=0} + I(A=0) \frac{\partial}{\partial t} h_{0t}(W,0,X)\big|_{t=0} \right \} \\
    =& \E\left \{ I(A=0) q_0(Z,X) \E\left[\frac{\partial}{\partial t} h_{0t}(W,1,X)\big|_{t=0} |Z,A=0,X\right] + I(A=0) \E\left[\frac{\partial}{\partial t} h_{0t}(W,0,X)\big|_{t=0} |Z,A=0,X\right] \right \} \\
    =& \E\left \{ I(A=0) q_0(Z,X) \E\left[ \frac{\partial}{\partial t} h_{1t}(W,M,1,X)\big|_{t=0} + \varepsilon_{01} S(W,M|Z,A=0,X) |Z,A=0,X\right] \right \} \\
    &+ \E\left \{ I(A=0) \E\left[ \frac{\partial}{\partial t} h_{1t}(W,M,0,X)\big|_{t=0} + \varepsilon_{00} S(W,M|Z,A=0,X) |Z,A=0,X\right] \right \} \quad (\text{Equation}~\eqref{eq: pathwise h0}) \\
    =& \E\left \{ I(A=0) q_0(Z,X) \frac{\partial}{\partial t} h_{1t}(W,M,1,X)\big|_{t=0} + I(A=0) q_0(Z,X) \varepsilon_{01} S(W,M|Z,A=0,X) \right \} \\
    &+ \E\left \{ I(A=0) \frac{\partial}{\partial t} h_{1t}(W,M,0,X)\big|_{t=0} + I(A=0) \varepsilon_{00} S(W,M|Z,A=0,X) \right \} \\
    =&: \E\left \{ (I)+(II) \right \} + \E\left \{ (III)+(IV) \right \}
\end{align*}}
Note that each term in the last equal sign can be respectively manipulated as below:
\begin{align*}
    &\E\left \{ (III) \right \} = \E\left \{ I(A=0) \frac{\partial}{\partial t} h_{1t}(W,M,0,X)\big|_{t=0} \right \} \\
    =& \E\left \{ I(A=0) \E\left[\frac{\partial}{\partial t} h_{1t}(W,M,A,X)\big|_{t=0} |Z,A,M,X\right] \right \} \\
    =& \E\left \{ I(A=0) \E\left [ \varepsilon_1 S(Y,W|Z,A,M,X) |Z,A,M,X \right ] \right \} \qquad (\text{Equation}~\eqref{eq: pathwise h1}) \\
    =& \E\left \{ I(A=0) \varepsilon_1 S(Y,W|Z,A,M,X) \right \} \\
    =& \E\left \{ I(A=0) \varepsilon_1 \left [ S(Y,W|Z,A,M,X) + S(Z,A,M,X) \right ] \right \} \\
    =& \E\left \{ I(A=0) \varepsilon_1 S(\mathcal{O}) \right \},
\end{align*}
where $\E\left \{ I(A=0) \varepsilon_1 S(Z,A,M,X) \right \} = 0$ follows from the fact that $\E(\varepsilon_1 |Z,A,M,X) = 0$, and
\begin{align*}
    &\E\left \{ (II) \right \} = \E\left \{ I(A=0) q_0(Z,X) \varepsilon_{01} S(W,M|Z,A=0,X) \right \} \\
    =& \E\left \{ I(A=0) q_0(Z,X) \varepsilon_{01} \left [ S(W,M|Z,A,X) + S(Z,A,X) \right ] \right \} \\
    =& \E\left \{ I(A=0) q_0(Z,X) \varepsilon_{01} \left [ S(W,M,Z,A,X) + S(Y|W,M,Z,A,X) \right ] \right \} \\
    =& \E\left \{ I(A=0) q_0(Z,X) \varepsilon_{01} S(\mathcal{O}) \right \},
\end{align*}
where the second equality follows from the observation that $\E\left \{ I(A=0) q_0(Z,X) \varepsilon_{01} S(Z,A,X) \right \} = \E\left \{ I(A=0) q_0(Z,X) S(Z,A,X) \E(\varepsilon_{01}|Z,A=0,X) \right \} = 0$ due to $\E(\varepsilon_{0a}|Z,A=0,X) = 0$, and
\begin{align*}
    &\E\left \{ (I) \right \} = \E\left \{ I(A=0) q_0(Z,X) \frac{\partial}{\partial t} h_{1t}(W,M,1,X)\big|_{t=0} \right \} \\
    =& \E\left \{ I(A=0) \E\left[q_0(Z,X) |W,A,M,X\right] \frac{\partial}{\partial t} h_{1t}(W,M,1,X)\big|_{t=0} \right \} \\
    =& \E\left \{ f(A=0|W,M,X) \E\left[q_0(Z,X) |W,A=0,M,X\right] \frac{\partial}{\partial t} h_{1t}(W,M,1,X)\big|_{t=0} \right \} \\
    =& \E\left \{ I(A=1) \frac{f(A=0|W,M,X)}{f(A=1|W,M,X)} \E\left[q_0(Z,X) |W,A=0,M,X\right] \frac{\partial}{\partial t} h_{1t}(W,M,1,X)\big|_{t=0} \right \} \\
    =& \E\left \{ I(A=1) \E\left[ q_1(Z,M,X)|W,A=1,M,X \right] \frac{\partial}{\partial t} h_{1t}(W,M,1,X)\big|_{t=0} \right \} \qquad (\text{Equation}~\eqref{eq: q1}) \\
    =& \E\left \{ I(A=1) q_1(Z,M,X) \frac{\partial}{\partial t} h_{1t}(W,M,1,X)\big|_{t=0} \right \} \\
    =& \E\left \{ I(A=1) q_1(Z,M,X) \E\left[ \frac{\partial}{\partial t} h_{1t}(W,M,A,X)\big|_{t=0} |Z,A,M,X\right] \right \} \\
    =& \E\left \{ I(A=1) q_1(Z,M,X) \E\left [ \varepsilon_1 S(Y,W|Z,A,M,X) |Z,A,M,X \right ] \right \} \qquad (\text{Equation}~\eqref{eq: pathwise h1}) \\
    =& \E\left \{ I(A=1) q_1(Z,M,X) \varepsilon_1 S(Y,W|Z,A,M,X) \right \} \\
    =& \E\left \{ I(A=1) q_1(Z,M,X) \varepsilon_1 \left [ S(Y,W|Z,A,M,X) + S(Z,A,M,X) \right ] \right \} \\
    =& \E\left \{ I(A=1) q_1(Z,M,X) \varepsilon_1 S(\mathcal{O}) \right \},
\end{align*}
where $\E\left \{ I(A=1) q_1(Z,M,X) \varepsilon_1 S(Z,A,M,X) \right \} = 0$ is implied by $\E(\varepsilon_1 |Z,A,M,X) = 0$, and
\begin{align*}
    &\E\left \{ (IV) \right \} = \E\left \{ I(A=0) \varepsilon_{00} S(W,M|Z,A=0,X) \right \} \\
    =& \E\left \{ I(A=0) \varepsilon_{00} \left [ S(W,M|Z,A,X) + S(Z,A,X) \right ] \right \} \\
    =& \E\left \{ I(A=0) \varepsilon_{00} \left [ S(W,M,Z,A,X) + S(Y|W,M,Z,A,X) \right ] \right \} \\
    =& \E\left \{ I(A=0) \varepsilon_{00} S(\mathcal{O}) \right \},
\end{align*}
where we employ $\E\left \{ I(A=0) \varepsilon_{00} S(Z,A,X) \right \} = \E\left \{ I(A=0) q_0(Z,X) S(Z,A,X) \E(\varepsilon_{01}|Z,A=0,X) \right \} = 0$ to obtain the second equality due to $\E(\varepsilon_{0a}|Z,A=0,X) = 0$.

Then, putting the above results together gives
\begin{align*}
    &\E\left [ \frac{\partial}{\partial t} h_{0t}(W,A,X)\big|_{t=0} \right ] \\
    =& \E\left \{ I(A=1) q_1(Z,M,X) \varepsilon_1 S(\mathcal{O}) + I(A=0) q_0(Z,X) \varepsilon_{01} S(\mathcal{O}) + I(A=0) \varepsilon_1 S(\mathcal{O}) +  I(A=0) \varepsilon_{00} S(\mathcal{O}) \right \}.
\end{align*}
Therefore,
\begin{align*}
    &I(A=1) q_1(Z,M,X) \varepsilon_1 + I(A=0) q_0(Z,X) \varepsilon_{01} + I(A=0) \varepsilon_1 + I(A=0) \varepsilon_{00} + h_{0}(W,A,X) - \Psi \\
    =& I(A=0) q_0(Z,X) \left [ h_1(W,M,1,X)-h_0(W,1,X) \right ] + I(A=0) \left [ h_1(W,M,0,X)-h_0(W,0,X) \right ] \\
    &+ \left [ I(A=1) q_1(Z, M, X) + I(A=0) \right ] \left [ Y-h_1(W,M,A,X) \right ] + h_0(W,A,X) - \Psi \\
    =&: EIF_{\Psi}
\end{align*}
is a valid influence function for the estimation of $\Psi$.

To show that $EIF_{\Psi} \in L_2(\mathcal{O})$ is the efficient influence function, it suffices to show that it belongs to the tangent space implied by the restrictions~\eqref{eq: pathwise h1} and \eqref{eq: pathwise h0} on the scores. Specifically, the tangent space is comprised of the set of mean zero scores $S(\mathcal{O}) \in L_2(\mathcal{O})$ satisfying that for $a = 0, 1$,
\begin{align*}
    \E\left [ \varepsilon_1 S(Y,W|Z,A,M,X) |Z,A,M,X \right ] &\in cl(R(T_1)), \\
    \E\left [ \varepsilon_{0a} S(W,M|Z,A=0,X) |Z,A=0,X \right ] &\in cl(R(T_0)),
\end{align*}
where $R()$ denotes the range space of an operator, and $cl()$ refers to the closure of a space. Evaluated at the submodel where Assumption~\ref{asm: T1 surjective} holds, $R(T_1)$ equals $L_2(Z,A,M,X)$ and $R(T_0)$ equals $L_2(Z,A=0,X)$, and thus the tangent space is essentially equal to $L_2(\mathcal{O})$ with zero mean, which completes the proof.

\subsection{Proof of Theorem~\ref{thm: multiply robust}}\label{app: proof of multiply robust}

We start with the proof of multiple robustness. Under some regularity conditions \citep{white1982maximum}, the nuisance estimators $\hat{\beta}_1, \hat{\beta}_0, \hat{\gamma}_1$ and $\hat{\gamma}_0$ would converge in probability to some population limits, denoted by $\beta^{\ast}_1, \beta^{\ast}_0, \gamma^{\ast}_1$ and $\gamma^{\ast}_0$, respectively.

Consider the semiparametric union model $\cM_{union} = \cM_1 \cup \cM_2 \cup \cM_3$. If under model $\cM_1$, i.e., $h_1(W,M,A,X;\beta^{\ast}_1)$ and $h_0(W,A,X;\beta^{\ast}_0)$ are correctly specified,
\begin{align*}
    \E(\hat{\Psi}_{P-MR}) =& \E\left\{ (1-A) q_0(Z,X; \gamma^{\ast}_0) \left [ h_1(W,M,1,X; \beta^{\ast}_1)-h_0(W,1,X; \beta^{\ast}_0) \right ] \right\} \\
    &+ \E \left \{ (1-A) \left [ h_1(W,M,0,X; \beta^{\ast}_1)-h_0(W,0,X; \beta^{\ast}_0) \right ] \right \} \\
    &+ \E \left \{ \left [ A q_1(Z,M,X; \gamma^{\ast}_1) + 1-A \right ] \left [ Y - h_1(W,M,A,X; \beta^{\ast}_1) \right ] \right \} + \E \left \{ h_0(W,A,X; \beta^{\ast}_0) \right \} \\
    =& \E\left\{ (1-A) q_0(Z,X; \gamma^{\ast}_0) \E\left [ h_1(W,M,1,X; \beta^{\ast}_1)-h_0(W,1,X; \beta^{\ast}_0) |Z,A=0,X \right ] \right\} \\
    &+ \E \left \{ (1-A) \E\left [ h_1(W,M,0,X; \beta^{\ast}_1)-h_0(W,0,X; \beta^{\ast}_0) |Z,A=0,X \right ] \right \} \\
    &+ \E \left \{ \left [ A q_1(Z,M,X; \gamma^{\ast}_1) + 1-A \right ] \E\left [ Y - h_1(W,M,A,X; \beta^{\ast}_1) |Z,A,M,X \right ] \right \} \\
    &+ \E \left \{ h_0(W,A,X; \beta^{\ast}_0) \right \} \\
    =& \Psi,
\end{align*}
implied by the integral equations~\eqref{eq: h1} and \eqref{eq: h0} and the identification~\eqref{eq: PIIE id}.

Next, if under model $\cM_2$, i.e., $h_1(W,M,A,X;\beta^{\ast}_1)$ and $q_0(Z,X;\gamma^{\ast}_0)$ are instead correctly specified,
\begin{align*}
    \E(\hat{\Psi}_{P-MR}) =& \E\left\{ (1-A) q_0(Z,X; \gamma^{\ast}_0) \left [ h_1(W,M,1,X; \beta^{\ast}_1)-h_0(W,1,X; \beta^{\ast}_0) \right ] \right\} \\
    &+ \E \left \{ (1-A) \left [ h_1(W,M,0,X; \beta^{\ast}_1)-h_0(W,0,X; \beta^{\ast}_0) \right ] \right \} + \E \left \{ h_0(W,A,X; \beta^{\ast}_0) \right \} \\
    &+ \E \left \{ \left [ A q_1(Z,M,X; \gamma^{\ast}_1) + 1-A \right ] \E\left [ Y - h_1(W,M,A,X; \beta^{\ast}_1) |Z,A,M,X \right ] \right \} \\
    =& \E \left \{ (1-A) q_0(Z,X; \gamma^{\ast}_0) h_1(W,M,1,X; \beta^{\ast}_1) + (1-A) h_1(W,M,0,X; \beta^{\ast}_1) \right \} \\
    &- \E\left\{ (1-A) q_0(Z,X; \gamma^{\ast}_0) h_0(W,1,X; \beta^{\ast}_0) + (1-A) h_0(W,0,X; \beta^{\ast}_0)\right\} + \E \left \{ h_0(W,A,X; \beta^{\ast}_0) \right \},
\end{align*}
implied by Equation~\eqref{eq: h1}. Note that by virtue of Equation~\eqref{eq: q0} and the identification~\eqref{eq: PIIE id-q0},
\begin{align*}
    \E\left\{ (1-A) q_0(Z,X; \gamma^{\ast}_0) h_0(W,1,X; \beta^{\ast}_0) \right\} =& \E\left\{ (1-A) h_0(W,1,X; \beta^{\ast}_0) \E\left[ q_0(Z,X; \gamma^{\ast}_0) |W,A=0,X \right] \right\} \\
    =& \E\left\{ (1-A) \frac{f(A=1|W,X)}{f(A=0|W,X)} h_0(W,1,X; \beta^{\ast}_0) \right\} \\
    =& \E\left\{ A h_0(W,1,X; \beta^{\ast}_0) \right\},
\end{align*}
and
\begin{equation*}
    \E \left \{ (1-A) q_0(Z,X; \gamma^{\ast}_0) h_1(W,M,1,X; \beta^{\ast}_1) + (1-A) h_1(W,M,0,X; \beta^{\ast}_1) \right \} = \Psi,
\end{equation*}
we have the consistency that $\E(\hat{\Psi}_{P-MR}) = \Psi$.

Finally, if under model $\cM_3$, i.e., $q_1(Z,M,X;\gamma^{\ast}_1)$ and $q_0(Z,X;\gamma^{\ast}_0)$ are instead correctly specified,
\begin{align*}
    \E(\hat{\Psi}_{P-MR}) =& \E\left\{ (1-A) q_0(Z,X; \gamma^{\ast}_0) \left [ h_1(W,M,1,X; \beta^{\ast}_1)-h_0(W,1,X; \beta^{\ast}_0) \right ] \right\} \\
    &+ \E \left \{ (1-A) \left [ h_1(W,M,0,X; \beta^{\ast}_1)-h_0(W,0,X; \beta^{\ast}_0) \right ] \right \} \\
    &+ \E \left \{ \left [ A q_1(Z,M,X; \gamma^{\ast}_1) + 1-A \right ] \left [ Y - h_1(W,M,A,X; \beta^{\ast}_1) \right ] \right \} + \E \left \{ h_0(W,A,X; \beta^{\ast}_0) \right \} \\
    =& \E \left \{ \left [ A q_1(Z,M,X; \gamma^{\ast}_1) + 1-A \right ] Y \right \} \\
    &+ \E \left \{ h_0(W,A,X; \beta^{\ast}_0) \right \} - \E\left\{ (1-A) q_0(Z,X; \gamma^{\ast}_0) h_0(W,1,X; \beta^{\ast}_0) \right\} \\
    &- \E \left \{ (1-A) h_0(W,0,X; \beta^{\ast}_0) \right \} \\
    &+ \E\left\{ \left[ (1-A) q_0(Z,X; \gamma^{\ast}_0) - A q_1(Z,M,X; \gamma^{\ast}_1) \right] h_1(W,M,1,X; \beta^{\ast}_1) \right\}.
\end{align*}
Note that the identification~\eqref{eq: PIIE id-q1} gives
\begin{equation*}
    \E \left \{ \left [ A q_1(Z,M,X; \gamma^{\ast}_1) + 1-A \right ] Y \right \} = \Psi,
\end{equation*}
it follows from Equation~\eqref{eq: q0} that
\begin{equation*}
    \E\left\{ (1-A) q_0(Z,X; \gamma^{\ast}_0) h_0(W,1,X; \beta^{\ast}_0) \right\} = \E\left\{ A h_0(W,1,X; \beta^{\ast}_0) \right\},
\end{equation*}
and Equation~\eqref{eq: q1} implies
{\small \begin{align*}
    &\E\left\{ \left[ (1-A) q_0(Z,X; \gamma^{\ast}_0) - A q_1(Z,M,X; \gamma^{\ast}_1) \right] h_1(W,M,1,X; \beta^{\ast}_1) \right\} \\
    =& \E\left\{ h_1(W,M,1,X; \beta^{\ast}_1) \left[ (1-A) \E\left(q_0(Z,X; \gamma^{\ast}_0) |W,M,A=0,X \right) - A \E\left(q_1(Z,M,X; \gamma^{\ast}_1) |W,M,A=1,X \right) \right] \right\} \\
    =& \E\left\{ h_1(W,M,1,X; \beta^{\ast}_1) \left[ f(A=0|W,M,X) \E\left(q_0(Z,X; \gamma^{\ast}_0) |W,M,A=0,X \right) \right.\right. \\
    &- \left.\left. f(A=1|W,M,X) \E\left(q_1(Z,M,X; \gamma^{\ast}_1) |W,M,A=1,X \right) \right] \right\} \\
    =& 0,
\end{align*}}
we obtain that $\E(\hat{\Psi}_{P-MR}) = \Psi$. Therefore, the multiple robustness of $\hat{\Psi}_{P-MR}$ follows.

In order to prove asymptotic normality and local efficiency, it suffices to show that $\hat{\Psi}_{P-MR}$ has an influence function equal to $EIF_{\Psi}$ under the intersection submodel model $\cM_{int} = \cM_1 \cap \cM_2 \cap \cM_3$. Following a Taylor expansion of $\hat{\Psi}_{P-MR}$ around $(\beta^{\ast}_1, \beta^{\ast}_0, \gamma^{\ast}_1, \gamma^{\ast}_0)$ and standard M-estimation arguments, we have that
\begin{align*}
    &\sqrt{n}(\hat{\Psi}_{P-MR} - \Psi) \\
    =& \frac{1}{\sqrt{n}} \sum_{i=1}^{n} \left\{ (1-A_i) q_0(Z_i,X_i; \gamma^{\ast}_0) \left [ h_1(W_i,M_i,1,X_i; \beta^{\ast}_1)-h_0(W_i,1,X_i; \beta^{\ast}_0) \right ] \right. \\
    &+ (1-A_i) \left [ h_1(W_i,M_i,0,X_i; \beta^{\ast}_1)-h_0(W_i,0,X_i; \beta^{\ast}_0) \right ] \\
    &+ \left. \left [ A_i q_1(Z_i,M_i,X_i; \gamma^{\ast}_1) + 1-A_i \right ] \left [ Y_i - h_1(W_i,M_i,A_i,X_i; \beta^{\ast}_1) \right ] + h_0(W_i,A_i,X_i; \beta^{\ast}_0) -\psi \right\} \\
    +& \sqrt{n} \left ( \hat{\gamma}_1 - \gamma^{\ast}_1 \right ) \E \left \{ A \frac{\partial q_1(Z,M,X; \gamma^{\ast}_1)}{\partial \gamma^{\ast}_1} \left [ Y - h_1(W,M,A,X; \beta^{\ast}_1) \right ] \right \} \\
    +& \sqrt{n} \left ( \hat{\gamma}_0 - \gamma^{\ast}_0 \right ) \E \left \{ (1-A) \frac{\partial q_0(Z,X; \gamma^{\ast}_0)}{\partial \gamma^{\ast}_0} \left [ h_1(W,M,1,X; \beta^{\ast}_1)-h_0(W,1,X; \beta^{\ast}_0) \right ] \right \} \\
    +& \sqrt{n} \left ( \hat{\beta}_1 - \beta^{\ast}_1 \right ) \E \left \{ \frac{\partial h_1(W,M,1,X; \beta^{\ast}_1)}{\partial \beta^{\ast}_1} \left [ (1-A) q_0(Z,X; \gamma^{\ast}_0) - A q_1(Z,M,X; \gamma^{\ast}_1) \right ] \right \} \\
    +& \sqrt{n} \left ( \hat{\beta}_0 - \beta^{\ast}_0 \right ) \E \left \{ \frac{\partial h_0(W,1,X; \beta^{\ast}_0)}{\partial \beta^{\ast}_0} \left [ A - (1-A) q_0(Z,X; \gamma^{\ast}_0) \right ] \right \} + o_p(1).
\end{align*}
According to the previous arguments in the proof of multiple robustness, it follows from Equations~\eqref{eq: h1}, \eqref{eq: h0}, \eqref{eq: q0_integral}, and \eqref{eq: q1_integral} that, under model $\cM_{int}$ where $h_1(W, M, A, X;\beta^{\ast}_1)$, $h_0(W, A, X;\beta^{\ast}_0)$, $q_1(Z, M, X;\gamma^{\ast}_1)$ and $q_0(Z, X;\gamma^{\ast}_0)$ are all correctly specified,
\begin{align*}
     \E \left \{ A \frac{\partial q_1(Z,M,X; \gamma^{\ast}_1)}{\partial \gamma^{\ast}_1} \left [ Y - h_1(W,M,A,X; \beta^{\ast}_1) \right ] \right \} =& 0, \\
     \E \left \{ (1-A) \frac{\partial q_0(Z,X; \gamma^{\ast}_0)}{\partial \gamma^{\ast}_0} \left [ h_1(W,M,1,X; \beta^{\ast}_1)-h_0(W,1,X; \beta^{\ast}_0) \right ] \right \} =& 0, \\
     \E \left \{ \frac{\partial h_0(W,1,X; \beta^{\ast}_0)}{\partial \beta^{\ast}_0} \left [ A - (1-A) q_0(Z,X; \gamma^{\ast}_0) \right ] \right \} =& 0, \\
     \E \left \{ \frac{\partial h_1(W,M,1,X; \beta^{\ast}_1)}{\partial \beta^{\ast}_1} \left [ (1-A) q_0(Z,X; \gamma^{\ast}_0) - A q_1(Z,M,X; \gamma^{\ast}_1) \right ] \right \} =& 0.
\end{align*}
These complete our proof.

\section{Debiased machine learning for proximal mediation inference}\label{sec: debiased ML}

We have proposed a semiparametric multiply robust estimator, $\hat{\Psi}_{P-MR}$, constructed using the efficient influence function for the target parameter, which initially relies on parametric estimation of intermediate nuisance functions. In this section, we extend our methodology to a more flexible framework that accommodates nonparametric estimation of nuisance functions. A key advantage of the multiply robust estimation based on the efficient influence function is its ability to seamlessly integrate modern machine learning techniques to handle high-dimensional nuisance parameters or functions, such as the confounding bridge functions $h_1$, $h_0$, $q_1$, and $q_0$, even when these nuisance estimators converge at rates slower than the standard $\sqrt{n}$-rate \citep{chernozhukov2018double, kennedy2024semiparametric, li2024identification}. Notably, applying our proposed proximal multiply robust estimator $\hat{\Psi}_{P-MR}$ within the debiased machine learning framework \citep{chernozhukov2018double} offers substantial gains in both efficiency and robustness. 

In high-dimension or highly complex settings, regularized machine learning approaches---such as lasso, ridge regression, boosting, or penalized neural networks---are often necessary for estimating nuisance functions. Regularization mitigates variance inflation but can also introduce bias into nuisance estimators. For instance, the convergence rate of the machine learning estimator $\hat{h}_1$ to $h_1$ in the root mean squared error sense is typically $n^{-\tau_{h_1}}$ with $\tau_{h_1} < 1/2$, which may prevent the naive plug-in estimator from achieving $\sqrt{n}$-consistency for $\Psi$. We will expound that our proposed proximal multiply robust estimation method can be adapted to debiased machine learning, ensuring $\sqrt{n}$-consistency even if the machine learning estimators of nuisance functions do not converge at the $\sqrt{n}$-rate in high-dimensional settings.

A fundamental property underpinning the validity of debiased machine learning is orthogonality, which ensures that the estimator for the parameter of interest remains robust to biases in nuisance function estimation. The estimator $\hat{\Psi}_{P-MR}$, as defined in Theorem~\ref{thm: multiply robust}, satisfies the estimation equation for $\Psi$:
\begin{equation}
    \frac{1}{n}\sum_{i=1}^{n} EIF_{\Psi}(\mathcal{O}_i; \Psi, \hat{\zeta}) = 0,
\label{eq: EIF estimating equation}
\end{equation}
where $EIF_{\Psi}$ is the efficient influence function given in Theorem~\ref{thm: efficiency bound}, $\mathcal{O}$ denotes the observed data $\mathcal{O} = \left ( Y, W, Z, A, M, X \right )$, and $\hat{\zeta}$ represents the estimator of nuisance bridge functions $\zeta = (h_1, h_0, q_1, q_0)$. 
Notably, the sensitivity of this estimating equation to errors in $\zeta$ is limited, as the Gateaux derivative with respect to $\zeta$ vanishes at the true nuisance functions $\zeta = (h_1, h_0, q_1, q_0)$. Using the notation from \cite{chernozhukov2018double}, the orthogonality condition can be expressed as:
\begin{equation*}
    \partial_{\zeta} \E \left\{ EIF_{\Psi}(\mathcal{O}_i; \Psi, \zeta) \right\} [\zeta^{\#} - \zeta] = 0, \quad \forall \zeta^{\#} \in \mathcal{T}_{\zeta}.
\end{equation*}
where $\mathcal{T}_{\zeta}$ represents the space of feasible nuisance function realizations. Specifically,
{\small \begin{align}
    \E\left \{ \frac{\partial EIF_{\Psi}(\mathcal{O}; \Psi, \zeta)}{\partial h_1} (h_1^{\#} - h_1) \right \} =& \E\left \{ \left [ (1-A) q_0 - A q_1 \right ] [h_1^{\#}(W,M,1,X) - h_1(W,M,1,X)] \right \} = 0, \label{eq: Neyman h1} \\
    \E\left \{ \frac{\partial EIF_{\Psi}(\mathcal{O}; \Psi, \zeta)}{\partial h_0} (h_0^{\#} - h_0) \right \} =& \E\left \{ \left [ A - (1-A) q_0 \right ] [h_0^{\#}(W,1,X) - h_0(W,1,X)] \right \} = 0, \label{eq: Neyman h0} \\
    \E\left \{ \frac{\partial EIF_{\Psi}(\mathcal{O}; \Psi, \zeta)}{\partial q_1} (q_1^{\#} - q_1) \right \} =& \E\left \{ A \left ( Y - h_1 \right ) (q_1^{\#} - q_1) \right \} = 0, \label{eq: Neyman q1} \\
    \E\left \{ \frac{\partial EIF_{\Psi}(\mathcal{O}; \Psi, \zeta)}{\partial q_0} (q_0^{\#} - q_0) \right \} =& \E\left \{  (1-A) \left [ h_1(W,M,1,X)-h_0(W,1,X) \right ] (q_0^{\#} - q_0) \right \} = 0, \label{eq: Neyman q0}
\end{align}}
for any possible bridge function realizations $(h_1^{\#}, h_0^{\#}, q_1^{\#}, q_0^{\#})$, as long as $h_1^{\#} \in L_2(W,M,A,X)$, $h_0^{\#} \in L_2(W,A,X)$, $q_1^{\#} \in L_2(Z,M,X)$, and $q_0^{\#} \in L_2(Z,X)$. The conditions~\eqref{eq: Neyman h1}-\eqref{eq: Neyman q0} are implied by integral equations~\eqref{eq: h1}, \eqref{eq: h0}, \eqref{eq: q0_integral}, and \eqref{eq: q1_integral}. This property, referred to as \textit{Neyman orthogonality} \citep{neyman1959optimal, neyman1979c}, ensures that the estimator remains insensitive to small perturbations in nuisance function estimation. Consequently, noisy estimates of nuisance functions can be used without strongly violating the moment condition of the estimating equation~\eqref{eq: EIF estimating equation}.

Beyond orthogonality, sample splitting plays a crucial role in mitigating bias due to overfitting, thereby ensuring the $\sqrt{n}$-consistency and asymptotic normality of debiased machine learning estimators. To clarify, suppose we randomly split the samples into two disjoint subsets: a main set of size $m$ with observations indexed by $i\in I$, and an auxiliary set of size $n-m$ with observations indexed by $i\in I^{c}$. The nuisance estimators $\hat{\zeta} = (\hat{h}_1, \hat{h}_0, \hat{q}_1, \hat{q}_0)$ are obtained using the auxiliary samples $i\in I^{c}$ via machine learning methods, and the final estimator $\hat{\Psi}_{P-MR}$ is then constructed using the main samples $i\in I$ based on nuisance estimates $\hat{\zeta}$:
\begin{align*}
       \frac{1}{m} \sum_{i\in I} &\left\{ (1-A_i) \hat{q}_0(Z_i,X_i) \left [ \hat{h}_1(W_i,M_i,1,X_i)-\hat{h}_0(W_i,1,X_i) \right ] \right. \\
       &+ (1-A_i) \left [ \hat{h}_1(W_i,M_i,0,X_i)-\hat{h}_0(W_i,0,X_i) \right ] \\
       &+ \left. \left [ A_i \hat{q}_1(Z_i,M_i,X_i) + 1-A_i \right ] \left [ Y_i - \hat{h}_1(W_i,M_i,A_i,X_i) \right ] + \hat{h}_0(W_i,A_i,X_i) \right\}.
\end{align*}
Without sample splitting, model errors, such as $(1-A_i) q_0(Z_i,X_i) - A_i q_1(Z_i,M_i,X_i)$, and estimation errors, such as $\hat{h}_1(W_i,M_i,1,X_i) - h_1(W_i,M_i,1,X_i)$, become generally correlated because the data $\mathcal{O}_i$ for observation $i$ is used in constructing the nuisance estimator $\hat{h}_1$. This overfitting issue can result in the term 
\begin{equation*}
    \frac{1}{\sqrt{n}} \sum_{i=1}^{n} \left [ (1-A_i) q_0(Z_i, X_i) - A_i q_1(Z_i, M_i, X_i) \right ] [\hat{h}_1(W_i,M_i,1,X_i) - h_1(W_i,M_i,1,X_i)]
\end{equation*}
failing to vanish asymptotically. Consequently, without sample splitting, even if $\hat{h}_1$ converges at a favorable rate (e.g., $n^{-1/2+\epsilon}$ for any $\epsilon >0$), the final estimator may still suffer from systematic bias of $\sqrt{n}$-order due to overfitting, thereby compromising its statistical efficiency.

While sample splitting mitigates overfitting bias, directly partitioning the dataset reduces efficiency since each subset is only partially used for estimating nuisance functions or constructing the final estimator. To overcome this limitation, we employ the cross-fitting technique \citep{schick1986asymptotically, chernozhukov2018double, ghassami2024causal}, which ensures that all observations contribute to both nuisance function estimation and parameter inference in separate folds. Specifically, we randomly divide the dataset into $L$ equal parts $\left \{ I_1, ..., I_L \right \}$. For each fold $l \in \left \{ 1, ..., L \right \}$, the nuisance functions are estimated using data excluding $I_l$, and the corresponding estimator for $\psi$ is constructed using these nuisance estimates and the observations in $I_l$. The final estimator for the parameter of interest is then obtained by averaging the $L$ fold-specific estimators. A key theoretical advantage of the cross-fitting procedure is that it relaxes the regularity conditions required for nuisance function estimation while still ensuring $\sqrt{n}$-consistency and asymptotic normality.

The following theorem establishes that, under mild regularity conditions, the proposed estimator remains $\sqrt{n}$-consistent and asymptotically normal, even when nuisance functions are estimated using machine learning methods that do not converge at the $\sqrt{n}$-rate. These machine learning approaches are employed to solve the integral equations~\eqref{eq: h1}, \eqref{eq: h0}, \eqref{eq: q0_integral}, and \eqref{eq: q1_integral}, which are constructed based on conditional mean-zero terms. Before stating the theorem, we impose a regularity assumption to ensure that these mean-zero terms have finite conditional second moments.

\begin{assumption}
    The mean-zero terms in integral equations~\eqref{eq: h1}, \eqref{eq: h0}, \eqref{eq: q0_integral}, and \eqref{eq: q1_integral}, possess bounded conditional second moments, i.e., for some constant $\eta >0$,
    \begin{align*}
        \E\left [ \left ( Y - h_1(W, M, A, X) \right )^2 \big| Z, A=1, M, X \right ] \leq& \eta, \\
        \E\left [ \left ( h_1(W, M, 1, X) - h_0(W, 1, X) \right )^2 \big| Z, A=0, X \right ] \leq& \eta, \\
        \E\left[ \left( (1-A) q_0(Z,X) -A \right)^2 |W, X \right] \leq& \eta, \\
        \E\left [ \left( A q_1(Z,M,X) - (1-A) q_0(Z,X) \right)^2 |W,M,X \right ] \leq& \eta.
    \end{align*}
\label{asm: finite second moments}
\end{assumption}

\begin{theorem}\emph{(\textbf{P-DML}):}
    Consider an $L$-fold random partition $\left \{ I_l \right \}_{l=1}^{L}$ of the observation indices $\left \{ 1, 2, ..., n \right \}$, such that each fold contains $m = n/L$ samples. For each fold $l$, construct machine learning estimators 
    \begin{equation*}
        \hat{\zeta}_l = \hat{\zeta}((\mathcal{O}_i)_{i\in I_l^{c}}) = (\hat{h}_{1_l}, \hat{h}_{0_l}, \hat{q}_{1_l}, \hat{q}_{0_l})
    \end{equation*}
    of $\zeta = (h_1, h_0, q_1, q_0)$ based on integral equations~\eqref{eq: h1}, \eqref{eq: h0}, \eqref{eq: q0_integral}, and \eqref{eq: q1_integral}, using observations indexed by $I_l^{c}$. Let $n^{-\tau_{h_1}}$, $n^{-\tau_{h_0}}$, $n^{-\tau_{q_1}}$, and $n^{-\tau_{q_0}}$ denote the respective convergence rates of $\hat{h}_{1_l}$ to $h_1$, $\hat{h}_{0_l}$ to $h_0$, $\hat{q}_{1_l}$ to $q_1$, and $\hat{q}_{0_l}$ to $q_0$ in the root mean square error sense, for some constants $\tau_{h_1}, \tau_{h_0}, \tau_{q_1}, \tau_{q_0} \geq 0$, such that
    \begin{equation}
        \tau_{h_1} + \tau_{q_1} \geq \frac{1}{2}, \quad \tau_{h_1} + \tau_{q_0} \geq \frac{1}{2}, \quad \tau_{h_0} + \tau_{q_0} \geq \frac{1}{2}. \label{limit: convergence rate}
    \end{equation}
    For each $l$, construct the estimator
    \begin{align*}
        \check{\Psi}_{(P-MR)_l} = \frac{1}{m} \sum_{i\in I_l} &\left\{ (1-A_i) \hat{q}_{0_l}(Z_i,X_i) \left [ \hat{h}_{1_l}(W_i,M_i,1,X_i)-\hat{h}_{0_l}(W_i,1,X_i) \right ] \right. \\
       &+ (1-A_i) \left [ \hat{h}_{1_l}(W_i,M_i,0,X_i)-\hat{h}_{0_l}(W_i,0,X_i) \right ] \\
       &+ \left. \left [ A_i \hat{q}_{1_l}(Z_i,M_i,X_i) + 1-A_i \right ] \left [ Y_i - \hat{h}_{1_l}(W_i,M_i,A_i,X_i) \right ] + \hat{h}_{0_l}(W_i,A_i,X_i) \right\}.
    \end{align*}
    Then, under Assumption~\ref{asm: finite second moments}, the aggregated estimator
    \begin{equation*}
        \check{\Psi}_{P-MR} = \frac{1}{L} \sum_{l=1}^{L} \check{\Psi}_{(P-MR)_l}
    \end{equation*}
    is a $\sqrt{n}$-consistent and asymptotically normal estimator of $\Psi$, satisfying
    \begin{equation*}
        \sqrt{n}(\check{\Psi}_{P-MR} - \Psi) = \frac{1}{\sqrt{n}} \sum_{i=1}^{n} EIF_{\Psi}(\mathcal{O}_i; \Psi, \zeta) + o_p(1).
    \end{equation*}
\label{thm: debiased ML}
\end{theorem}

A notable special case in which condition~\eqref{limit: convergence rate} holds is when $\tau_{h_1}, \tau_{h_0}, \tau_{q_1}, \tau_{q_0} = 1/4$, implying that the estimators $\hat{h}_1$, $\hat{h}_0$, $\hat{q}_1$, and $\hat{q}_0$ are only required to achieve an $o\left(n^{-1/4}\right)$ convergence rate in the root mean square error sense. In contrast, parametric models for nuisance parameters typically attain an $O\left ( 1/\sqrt{n} \right )$ convergence rate. Therefore, Theorem~\ref{thm: debiased ML} accommodates scenarios where $\hat{h}_1$, $\hat{h}_0$, $\hat{q}_1$, and $\hat{q}_0$ converge at a rate significantly slower than the parametric benchmark. The $o\left(n^{-1/4}\right)$ convergence rate is achievable for various machine learning methods \citep{chen1999improved, Belloni2013least} under appropriate structural assumptions on nuisance parameters. Notably, \cite{kallus2021causal} and \cite{ghassami2022minimax} introduced a kernel-based estimation approach for semiparametric proximal causal inference, leveraging nonparametric adversarial learning techniques for solving conditional moment equations \citep{dikkala2020minimax}. In Section~\ref{app: minimax bridge}, we extend this methodology to proximal mediation analysis, presenting a minimax learning approach for estimating bridge functions that satisfy integral equations~\eqref{eq: h1}, \eqref{eq: h0}, \eqref{eq: q0_integral} and \eqref{eq: q1_integral}.

\subsection{Proof of Theorem~\ref{thm: debiased ML}}

For each $l \in \left \{ 1, 2,...,L \right \}$, let $n^{-\tau_{h_1}}$, $n^{-\tau_{h_0}}$, $n^{-\tau_{q_1}}$, and $n^{-\tau_{q_0}}$ represent respectively the convergence rates of $\hat{h}_{1_l}$ to $h_1$, $\hat{h}_{0_l}$ to $h_0$, $\hat{q}_{1_l}$ to $q_1$, and $\hat{q}_{0_l}$ to $q_0$ in the root mean square error sense, indicating that
\begin{align}
    \sqrt{\E\left [ \left | \hat{h}_{1_l}(W,M,A,X) - h_1(W,M,A,X) \right |^2 \right ]} =& o\left ( n^{-\tau_{h_1}} \right ), \label{limit: h1} \\
    \sqrt{\E\left [ \left | \hat{h}_{0_l}(W,A,X) - h_0(W,A,X) \right |^2 \right ]} =& o\left ( n^{-\tau_{h_0}} \right ), \label{limit: h0} \\
    \sqrt{\E\left [ \left | \hat{q}_{1_l}(Z,M,X) - q_1(Z,M,X) \right |^2 \right ]} =& o\left ( n^{-\tau_{q_1}} \right ), \label{limit: q1} \\
    \sqrt{\E\left [ \left | \hat{q}_{0_l}(Z,X) - q_0(Z,X) \right |^2 \right ]} =& o\left ( n^{-\tau_{q_0}} \right ). \label{limit: q0}
\end{align}

For the $l$-th fold random partition, we contrast $\check{\Psi}_{(P-MR)_l}$ with the oracle estimator $\Psi_{(P-MR)_l}$ that takes the same form as $\check{\Psi}_{(P-MR)_l}$ but replaces the machine learning estimators $\hat{\zeta}_l$ by oracle ones $\zeta$. After some simple arithmetic, we obtain the decomposition as follows:
\begin{align*}
    &\sqrt{n}\left ( \check{\Psi}_{(P-MR)_l} - \Psi_{(P-MR)_l} \right ) \\
    =& \sqrt{n} \frac{1}{m} \sum_{i\in I_l} \left\{ (1-A_i) \hat{q}_{0_l}(Z_i,X_i) \left [ \hat{h}_{1_l}(W_i,M_i,1,X_i)-\hat{h}_{0_l}(W_i,1,X_i) \right ] \right. \\
    &\quad + (1-A_i) \left [ \hat{h}_{1_l}(W_i,M_i,0,X_i)-\hat{h}_{0_l}(W_i,0,X_i) \right ] \\
    &\quad + \left. \left [ A_i \hat{q}_{1_l}(Z_i,M_i,X_i) + 1-A_i \right ] \left [ Y_i - \hat{h}_{1_l}(W_i,M_i,A_i,X_i) \right ] + \hat{h}_{0_l}(W_i,A_i,X_i) \right\} \\
    &- \sqrt{n} \frac{1}{m} \sum_{i\in I_l} \left\{ (1-A_i) q_{0}(Z_i,X_i) \left [ h_{1}(W_i,M_i,1,X_i)-h_{0}(W_i,1,X_i) \right ] \right. \\
    &\quad + (1-A_i) \left [ h_{1}(W_i,M_i,0,X_i) - h_{0}(W_i,0,X_i) \right ] \\
    &\quad + \left. \left [ A_i q_{1}(Z_i,M_i,X_i) + 1-A_i \right ] \left [ Y_i - h_{1}(W_i,M_i,A_i,X_i) \right ] + h_{0}(W_i,A_i,X_i) \right\} \\
    =& \sqrt{n} \frac{1}{m} \sum_{i\in I_l} A_i \left [ Y_i - h_{1}(W_i,M_i,A_i,X_i) \right ] \left [ \hat{q}_{1_l}(Z_i,M_i,X_i) - q_{1}(Z_i,M_i,X_i) \right ] \\
    &+ \sqrt{n} \frac{1}{m} \sum_{i\in I_l} (1-A_i) \left [ h_{1}(W_i,M_i,1,X_i)-h_{0}(W_i,1,X_i) \right ] \left [ \hat{q}_{0_l}(Z_i,X_i) - q_{0}(Z_i,X_i) \right ] \\
    &+ \sqrt{n} \frac{1}{m} \sum_{i\in I_l} \left [ (1-A_i) q_0(Z_i, X_i) - A_i q_1(Z_i, M_i, X_i) \right ] \left [ \hat{h}_{1_l}(W_i, M_i, 1, X_i) - h_1(W_i, M_i, 1, X_i) \right ] \\
    &+ \sqrt{n} \frac{1}{m} \sum_{i\in I_l} \left [ A_i - (1-A_i) q_0(Z_i, X_i) \right ] \left [ \hat{h}_{0_l}(W_i, 1, X_i) - h_0(W_i, 1, X_i) \right ] \\
    &- \sqrt{n} \frac{1}{m} \sum_{i\in I_l} A_i \left [ \hat{q}_{1_l}(Z_i,M_i,X_i) - q_{1}(Z_i,M_i,X_i) \right ] \left [ \hat{h}_{1_l}(W_i, M_i, 1, X_i) - h_1(W_i, M_i, 1, X_i) \right ] \\
    &+ \sqrt{n} \frac{1}{m} \sum_{i\in I_l} (1-A_i) \left [  \hat{q}_{0_l}(Z_i,X_i) - q_{0}(Z_i,X_i) \right ] \left [ \hat{h}_{1_l}(W_i, M_i, 1, X_i) - h_1(W_i, M_i, 1, X_i) \right ] \\
    &- \sqrt{n} \frac{1}{m} \sum_{i\in I_l} (1-A_i) \left [  \hat{q}_{0_l}(Z_i,X_i) - q_{0}(Z_i,X_i) \right ] \left [ \hat{h}_{0_l}(W_i, 1, X_i) - h_0(W_i, 1, X_i) \right ] \\
    =:& (I) + (II) + (III) +(IV) - (V) + (VI) - (VII).
\end{align*}
We will then verify that these terms are small asymptotically.

For terms $(I)$-$(IV)$, thanks to the cross-fitting construction, nuisance function estimators $\hat{q}_{1_l}$, $\hat{q}_{0_l}$, $\hat{h}_{1_l}$, and $\hat{h}_{0_l}$, are deterministic when conditional on data from $I_{l}^{c}$. Further, conditionally on $I_{l}^{c}$, the first four terms are mean-zero implied by the integral equations~\eqref{eq: h1}-\eqref{eq: h0} and \eqref{eq: q0_integral}-\eqref{eq: q1_integral} that define bridge functions. Additionally, they have asymptotically small variances under Assumption~\ref{asm: finite second moments}. Take $(I)$ as an example:
{\small \begin{align*}
    &Var\left\{ \sqrt{n} \frac{1}{m} \sum_{i\in I_l} A_i \left [ Y_i - h_{1}(W_i,M_i,A_i,X_i) \right ] \left [ \hat{q}_{1_l}(Z_i,M_i,X_i) - q_{1}(Z_i,M_i,X_i) \right ] \big| \left \{ Z_i, M_i, A_i, X_i \right \}_{i\in I_l}, \left \{ \mathcal{O}_j \right \}_{j\in I_{l}^{c}} \right\} \\
    =& \frac{n}{m^2} \sum_{i\in I_l} \E\left \{ A_i \left [ Y_i - h_{1}(W_i,M_i,A_i,X_i) \right ]^2 \left [ \hat{q}_{1_l}(Z_i,M_i,X_i) - q_{1}(Z_i,M_i,X_i) \right ]^2 \big| Z_i, M_i, A_i, X_i, \left \{ \mathcal{O}_j \right \}_{j\in I_{l}^{c}} \right \} \\
    =& \frac{L}{m} \sum_{i\in I_l} A_i \left [ \hat{q}_{1_l}(Z_i,M_i,X_i) - q_{1}(Z_i,M_i,X_i) \right ]^2 \E\left \{ \left [ Y_i - h_{1}(W_i,M_i,A_i,X_i) \right ]^2  \big| Z_i, M_i, A_i=1, X_i, \left \{ \mathcal{O}_j \right \}_{j\in I_{l}^{c}} \right \} \\
    \leq& \frac{L \eta}{m} \sum_{i\in I_l} \left [ \hat{q}_{1_l}(Z_i,M_i,X_i) - q_{1}(Z_i,M_i,X_i) \right ]^2
    =o_p\left ( \frac{1}{n^{2\tau_{q_1}}} \right ),
\end{align*}}
where the inequality is due to Assumption~\ref{asm: finite second moments}, and the last equality sign is implied by \eqref{limit: q1}. Hence, note that $\tau_{q_1} \geq 0$, it follows from Chebyshev's inequality that $(I)$ converges to zero in probability. Furthermore, under Assumption~\ref{asm: finite second moments}, the terms $(II)$, $(III)$, and $(IV)$ converge to zero in probability by a similar argument.

For terms $(V)$-$(VII)$, we take $(V)$ as an example:
\begin{align*}
    &\sqrt{n} \frac{1}{m} \sum_{i\in I_l} A_i \left [ \hat{q}_{1_l}(Z_i,M_i,X_i) - q_{1}(Z_i,M_i,X_i) \right ] \left [ \hat{h}_{1_l}(W_i, M_i, 1, X_i) - h_1(W_i, M_i, 1, X_i) \right ] \\
    \leq& \sqrt{n} \sqrt{\frac{1}{m} \sum_{i\in I_l} \left [ \hat{q}_{1_l}(Z_i,M_i,X_i) - q_{1}(Z_i,M_i,X_i) \right ]^2} \sqrt{\frac{1}{m} \sum_{i\in I_l} \left [ \hat{h}_{1_l}(W_i, M_i, 1, X_i) - h_1(W_i, M_i, 1, X_i) \right ]^2} \\
    =& o_p\left ( \frac{\sqrt{n}}{n^{\tau_{q_1} + \tau_{h_1}}} \right ),
\end{align*}
where the inequality sign results from the Cauchy-Schwarz inequality, and the last equality sign is implied by \eqref{limit: h1} and \eqref{limit: q1}. Thus, note that $\tau_{q_1} + \tau_{h_1} \geq \frac{1}{2}$, we have that $(V)$ converges to zero in probability. Furthermore, following a similar argument, we find that $(VI) \leq o_p\left ( \frac{\sqrt{n}}{n^{\tau_{q_0} + \tau_{h_1}}} \right )$ and $(VII) \leq o_p\left ( \frac{\sqrt{n}}{n^{\tau_{q_0} + \tau_{h_0}}} \right )$. Hence, because $\tau_{q_0} + \tau_{h_1} \geq \frac{1}{2}$ and $\tau_{q_0} + \tau_{h_0} \geq \frac{1}{2}$, we obtain that the terms $(VI)$ and $(VII)$ also converge to zero in probability.

Therefore, we get that
\begin{equation*}
    \sqrt{n}\left ( \check{\Psi}_{(P-MR)_l} - \Psi_{(P-MR)_l} \right ) \stackrel{p}{\longrightarrow} 0.
\end{equation*}
Note that $\frac{1}{n} \sum_{i=1}^{n} EIF_{\Psi}(\mathcal{O}_i; \Psi, \zeta) = \frac{1}{L} \sum_{l=1}^{L} \Psi_{(P-MR)_l} -\Psi$, the $\sqrt{n}-$consistency and normality for $\check{\Psi}_{P-MR}$ can then be proven by carrying out the same argument for other folds to obtain
\begin{equation*}
    \sqrt{n}(\check{\Psi}_{P-MR} - \Psi) = \frac{1}{\sqrt{n}} \sum_{i=1}^{n} EIF_{\Psi}(\mathcal{O}_i; \Psi, \zeta) + o_p(1).
\end{equation*}

\subsection{Minimax estimation for bridge functions}\label{app: minimax bridge}

We proposed a debiased machine learning estimation approach leveraging the efficient influence function, which remains $\sqrt{n}$-consistent even when machine learning-based nuisance function estimators do not converge at the $\sqrt{n}$-rate. This section focuses on estimating nuisance functions using a recently developed minimax learning approach.

The nuisance bridge functions $h_1$, $h_0$, $q_1$, and $q_0$ are solutions to integral equations~\eqref{eq: h1}, \eqref{eq: h0}, \eqref{eq: q0}, and \eqref{eq: q1}, precluding direct application of standard regression techniques. Instead, we adopt a nonparametric adversarial learning method for solving these conditional moment equations, following \cite{dikkala2020minimax} and its adaptations for the original semiparametric proximal causal inference in \cite{kallus2021causal} and
\cite{ghassami2022minimax}. Using this approach, we estimate the nuisance functions $h_1$, $h_0$, $q_0$, and $q_1$ via a minimax kernel method. Let $\cH_1$, $\cH_{0}$, $\mathcal{Q}_0$, $\mathcal{Q}_1$, and $\cG$ denote normed function spaces. 

For bridge functions $h_1$ and $h_0$, which solve conditional moment equations~\eqref{eq: h1} and \eqref{eq: h0}, we propose the nonparametric regularized estimators based on the following optimizations:

{\footnotesize 
\begin{align*}
    &\hat{h}_1(W,M,A,X) = \arg\min_{h_1 \in \cH_1} \sup_{g \in \mathcal{G}} \E_n\left \{ \left [ Y-h_1(W,M,A,X) \right ] g(Z,A,M,X) - g^2(Z,A,M,X) \right \} - \lambda_{\mathcal{G}}^{h_1} \left \| g \right \|_{\cG}^2 + \lambda_{\cH_1}^{h_1} \left \| h_1 \right \|_{\cH_1}^2, \\
    &\hat{h}_0(W,1,X) = \arg\min_{h_{0} \in \cH_{0}} \sup_{g \in \mathcal{G}} \E_{n_0}\left \{ \left [ \hat{h}_1(W,M,1,X) - h_{0}(W,X) \right ] g(Z,X) - g^2(Z,X) \right \} - \lambda_{\mathcal{G}}^{h_{01}} \left \| g \right \|_{\cG}^2 + \lambda_{\cH_{01}}^{h_{01}} \left \| h_{0} \right \|_{\cH_{01}}^2, \\
    &\hat{h}_0(W,0,X) = \arg\min_{h_{0} \in \cH_{0}} \sup_{g \in \mathcal{G}} \E_{n_0}\left \{ \left [ \hat{h}_1(W,M,0,X) - h_{0}(W,X) \right ] g(Z,X) - g^2(Z,X) \right \} - \lambda_{\mathcal{G}}^{h_{00}} \left \| g \right \|_{\cG}^2 + \lambda_{\cH_{00}}^{h_{00}} \left \| h_{0} \right \|_{\cH_{00}}^2,
\end{align*}
}
where $\E_n(.)$ and $\E_{n_0}(.)$ represent empirical expectations over the full dataset and the subset with $A=0$, respectively. For bridge functions $q_0$ and $q_1$, which solve conditional moment equations~\eqref{eq: q0_integral} and \eqref{eq: q1_integral}, we propose the nonparametric regularized estimators based on the optimizations as follows:

{\footnotesize 
\begin{align*}
    &\hat{q}_0(Z,X) = \arg\min_{q_0 \in \mathcal{Q}_0} \sup_{g \in \mathcal{G}} \E_n\left \{ \left [ (1-A)q_0(Z,X) - A \right ] g(W,X) - g^2(W,X) \right \} - \lambda_{\mathcal{G}}^{q_0} \left \| g \right \|_{\cG}^2 + \lambda_{\mathcal{Q}_0}^{q_0} \left \| q_0 \right \|_{\mathcal{Q}_0}^2, \\
    &\hat{q}_1(Z,M,X) = \arg\min_{q_1 \in \mathcal{Q}_1} \sup_{g \in \mathcal{G}} \E_n\left \{ \left [ A q_1(Z,M,X) - (1-A)\hat{q}_0(Z,X) \right ] g(W,M,X) - g^2(W,M,X) \right \} - \lambda_{\mathcal{G}}^{q_1} \left \| g \right \|_{\cG}^2 + \lambda_{\mathcal{Q}_1}^{q_1} \left \| q_1 \right \|_{\mathcal{Q}_1}^2.
\end{align*}
}

The convergence properties of the minimax estimators are detailed in \cite{dikkala2020minimax} and \cite{ghassami2022minimax}. Here, we provide an intuitive explanation for these minimax estimators. Take $\hat{h}_1$ as an example, with similar arguments extending to $\hat{h}_0$, $\hat{q}_0$, and $\hat{q}_1$. Note that
\begin{align*}
    &\arg\sup_{g \in \mathcal{G}} \E\left \{ \left [ Y-h_1(W,M,A,X) \right ] g(Z,A,M,X) - g^2(Z,A,M,X) \right \} \\
    =& \arg\sup_{g \in \mathcal{G}} \E\left \{ \left [ Y-h_1(W,M,A,X) \right ] g(Z,A,M,X) - g^2(Z,A,M,X) \right. \\
    &- \left. \frac{1}{4}\left [ Y-h_1(W,M,A,X) \right ]^2 + \frac{1}{4}\left [ Y-h_1(W,M,A,X) \right ]^2 \right \} \\
    =& \arg\sup_{g \in \mathcal{G}} \E\left \{ \frac{1}{4}\left [ Y-h_1(W,M,A,X) \right ]^2 -\left ( g(Z,A,M,X) - \frac{1}{2}\left [ Y-h_1(W,M,A,X) \right ] \right )^2 \right \} \\
    =& \arg\inf_{g \in \mathcal{G}} \E\left \{ \left ( g(Z,A,M,X) - \frac{1}{2}\left [ Y-h_1(W,M,A,X) \right ] \right )^2 \right \} \\
    =& \frac{1}{2} \E\left \{ \left [ Y-h_1(W,M,A,X) \right ] \Big|Z,A,M,X \right \}.
\end{align*}
It follows that 
{\small \begin{align*}
    &\sup_{g \in \mathcal{G}} \E\left \{ \left [ Y-h_1(W,M,A,X) \right ] g(Z,A,M,X) - g^2(Z,A,M,X) \right \} \\
    =& \E\left \{ \frac{1}{2} \left [ Y-h_1(W,M,A,X) \right ] \E \left [ Y-h_1(W,M,A,X) \Big|Z,A,M,X \right ] - \frac{1}{4} \E \left [ Y-h_1(W,M,A,X) \Big|Z,A,M,X \right ]^2 \right \} \\
    =& \frac{1}{4} \E\left \{ \E \left [ Y-h_1(W,M,A,X) \Big|Z,A,M,X \right ]^2 \right \},
\end{align*}}
which is always non-negative, and achieves zero if and only if $h_1$ satisfies the integral equation~\eqref{eq: h1}. Consequently, minimizing the empirical counterpart of this expression with respect to $h_1$ yields the minimax estimator $\hat{h}_1$. The inclusion of regularization terms prevents overfitting and ensures favorable convergence rates \citep{dikkala2020minimax, ghassami2022minimax}.

\section{Supplementary to numerical experiments}\label{app: numerical experiments}

In Section~\ref{app: DGP}, we present the data-generating process in the numerical experiments. In Section~\ref{app: Choices of bridge functions for numerical experiments}, we specify and justify the models for bridge functions that satisfy the estimating equations~\eqref{eq: h1}-\eqref{eq: q1} under this data-generating process.

\subsection{Data-generating process}\label{app: DGP}

First, the covariates $X = (X_1, X_2)^T$ and the unobserved confounder $U$ are drawn from a multivariate normal distribution, $(X_1, X_2, U)^T \sim \cN(\mu, \Sigma)$, where $\mu = (0.25, 0.25, 0)^{T}$ and \begin{equation*}
    \Sigma = \begin{pmatrix}
             \sigma_{x_1}^2 & \sigma_{x_1 x_2} & \sigma_{x_1 u} \\
             \sigma_{x_1 x_2} & \sigma_{x_2}^2 & \sigma_{x_2 u} \\
             \sigma_{x_1 u} & \sigma_{x_2 u} & \sigma_{u}^2
             \end{pmatrix}
           = \begin{pmatrix}
             0.25 & 0 & 0.05 \\
             0 & 0.25 & 0.05 \\
             0.05 & 0.05 & 1
             \end{pmatrix}.
\end{equation*} 
Conditional on $(X, U)$, the exposure $A$ is generated from a Bernoulli distribution with $f(A=1|X,U) = \expit(-(0.5, 0.5)X - 0.4U)$. The proxy variables are generated according to $Z|A,X,U \sim \cN(0.2-0.52A+(0.2,0.2)X-U, 1)$ and $W|X,U \sim \cN(0.3+(0.2,0.2)X-0.6U, 1)$. We define the intervening variable deterministically as $A_M = A$, and generate the mediator $M$ conditional on $(A_M, X, U)$ such that $M|A_M,X,U \sim \cN(-0.3A_M - (0.5,0.5)X+1.5U, 1)$. Finally, the outcome $Y$ is generated as: $$Y = 2+2A+M+2W-(1,1)X-U+2\epsilon^{\ast}, \quad \epsilon^{\ast} \sim \cN(0,1).$$ 

As shown in the supplementary material, the above data-generating mechanism is compatible with the following models for the outcome confounding bridge functions:
\begin{align}
    h_1(W,M,A,X;\beta_1) =& \beta_{1,0} + \beta_{1,w} W + \beta_{1,m} M + \beta_{1,a} A + \beta_{1,x}^{T} X, \label{eq: h1_simu} \\
    h_0(W,A,X;\beta_0) =& \beta_{0,0} + \beta_{0,w} W + \beta_{0,a} A + \beta_{0,x}^{T} X, \label{eq: h0_simu}
\end{align}
respectively, such that the choice of $h_1(W,M,A,X;\beta_1)$ and $ h_0(W,A,X;\beta_0)$ satisfies the integral equations~\eqref{eq: h1} and \eqref{eq: h0}. Furthermore, the supplementary material elaborates that when $A$ follows a Bernoulli distribution given $U$ and $X$, $M$ is normally distributed given $U$, $A$ and $X$, and $Z \sim \cN(\epsilon_0 + \epsilon_u U + \epsilon_a A + \epsilon_x X, \sigma_{z|u, a, x}^2)$ given $U$, $A$ and $X$, then the choice of exposure confounding bridge functions:
\begin{align}
    q_0(Z,X) =& \exp\left \{ -(\gamma_{0,0} + \gamma_{0,z}Z + \gamma_{0,x}X) \right \}, \label{eq: q0_simu} \\
    q_1(Z,M,X) =& q_0(Z,X) \exp\left \{ \gamma_{1,0} + \gamma_{1,z}Z + \gamma_{1,m}M + \gamma_{1,x}X \right \}, \label{eq: q1_simu}
\end{align}
satisfies the integral equations~\eqref{eq: q0} and \eqref{eq: q1}.

\subsection{Choices of bridge functions for numerical experiments}\label{app: Choices of bridge functions for numerical experiments}

We first elaborate that under the data-generating mechanism in simulation studies, outcome confounding bridge functions have the forms~\eqref{eq: h1_simu} and \eqref{eq: h0_simu}.
It suffices to verify that expressions~\eqref{eq: h1_simu} and \eqref{eq: h0_simu} satisfy integral equations~\eqref{eq: h1} and \eqref{eq: h0}. Note that $Y$ is generated from $Y = 2+2A+M+2W-(1,1)^{T}X-U+2\epsilon^{\ast}$, and $W \indep \left \{ A, M \right \}|U, X$, we have that

\begin{align*}
    \E(Y|U,A,M,X) =& 2 + 2A + M + 2\E(W|U,A,M,X) - (1,1)^{T}X - U \\
    =& 2 + 2A + M + 2\E(W|U,X) - (1,1)^{T}X - U \\
    =& 2.6 + 2A + M -(0.6,0.6)^{T} X - 2.2U.
\end{align*}

For the expression~\eqref{eq: h1_simu} that $h_1(W,M,A,X;\beta_1) = \beta_{1,0} + \beta_{1,w} W + \beta_{1,m} M + \beta_{1,a} A + \beta_{1,x}^{T} X$,
{\small \begin{align*}
    \E[ h_1(W,M,A,X;\beta_1) |U,A,M,X] =& \beta_{1,0} + \beta_{1,w} \E(W|U,A,M,X) + \beta_{1,m} M + \beta_{1,a} A + \beta_{1,x}^{T} X \\
    =& \beta_{1,0} + \beta_{1,w} \E(W|U,X) + \beta_{1,m} M + \beta_{1,a} A + \beta_{1,x}^{T} X \\
    =& (\beta_{1,0} + 0.3\beta_{1,w}) + \beta_{1,m} M + \beta_{1,a} A + (\beta_{1,x}^{T} + \beta_{1,w}(0.2,0.2)^{T}) X - 0.6\beta_{1,w} U.
\end{align*}}
After matching the coefficients, it follows that
\begin{equation*}
    \left\{\begin{matrix}
     \beta_{1,0} = 1.5 \\
     \beta_{1,m} = 1 \\
     \beta_{1,a} = 2 \\
     \beta_{1,x} = -(\frac{4}{3}, \frac{4}{3}) \\
     \beta_{1,w} = \frac{11}{3}
    \end{matrix}\right.
\end{equation*}
are compatible with the integral equation~\eqref{eq: h1}. In other words,
\begin{equation*}
    h_1(W,M,A,X) = 1.5 + \frac{11}{3} W + M +2A -\left(\frac{4}{3}, \frac{4}{3}\right)^{T} X
\end{equation*}
is a proper outcome confounding bridge function satisfying \eqref{eq: h1}.

Then, it turns out to be
\begin{align*}
    \E\left [ h_1(W,M,a,X) |U, A=0, X \right ] =& 1.5 + \frac{11}{3} \E(W|U, A=0, X) + \E(M|U, A=0, X) + 2a -\left(\frac{4}{3}, \frac{4}{3}\right)^{T} X \\
    =& 1.5 + \frac{11}{3} \E(W|U, X) + \E(M|U, A=0, X) + 2a -\left(\frac{4}{3}, \frac{4}{3}\right)^{T} X \\
    =& 2.6 - (1, 1)^{T} X - 0.7U + 2a,
\end{align*}
for $a = 0, 1$. Consider the expression~\eqref{eq: h0_simu} that $h_0(W,A,X;\beta_0) = \beta_{0,0} + \beta_{0,w} W + \beta_{0,a} A + \beta_{0,x}^{T} X$,
\begin{align*}
    \E\left [ h_0(W,a,X;\beta_0) |U, A=0, X \right ] =& \beta_{0,0} + \beta_{0,w} \E(W|U, A=0, X) + \beta_{0,a} a + \beta_{0,x}^{T} X \\
    =& \beta_{0,0} + \beta_{0,w} \E(W|U, X) + \beta_{0,a} a + \beta_{0,x}^{T} X \\
    =& \beta_{0,0} + 0.3\beta_{0,w} + (\beta_{0,x}^{T} + \beta_{0,w}(0.2, 0.2)^{T}) X -0.6\beta_{0,w} U + \beta_{0,a} a,
\end{align*}
for $a = 0, 1$. After matching the coefficients, it follows that
\begin{equation*}
    \left\{\begin{matrix}
     \beta_{0,0} = 2.25 \\
     \beta_{0,a} = 2 \\
     \beta_{0,x} = -(\frac{4}{3}, \frac{4}{3}) \\
     \beta_{0,w} = \frac{7}{6}
    \end{matrix}\right.
\end{equation*}
are compatible with the integral equation~\eqref{eq: h0}. In other words,
\begin{equation*}
    h_0(W,A,X) = 2.25 + \frac{7}{6} W + 2A -\left(\frac{4}{3}, \frac{4}{3}\right)^{T} X
\end{equation*}
is a proper outcome confounding bridge function satisfying \eqref{eq: h0}. Therefore, the linear models specified in \eqref{eq: h1_simu} and \eqref{eq: h0_simu} are the outcome confounding bridge functions $h_1$ and $h_0$ in our data-generating mechanism.

Next, we consider the specification of the exposure confounding bridge functions, $q_0$ and $q_1$, under the data-generating mechanism described in simulation studies. For ease of exposition, we denote the exposure confounding bridge functions introduced in \cite{dukes2023proximal} as $q_0^{\ast}$ and $q_1^{\ast}$. We observe the connection between $q_0$ and $q_0^{\ast}$ that $$q_0(Z, X) = q_0^{\ast}(Z, X) -1.$$ Moreover, $q_1$ and $q_1^{\ast}$ satisfy the same integral equations. Consequently, we adapt the results of \cite{dukes2023proximal} to specify the exposure confounding bridge functions $q_0$ and $q_1$ in our setting, as formalized in the following corollary.

\begin{corollary}
    When $A$ follows a Bernoulli distribution given $U$ and $X$ with $\frac{1}{f(A=0|U,X)} = 1 + \exp\left \{ -(\alpha_{0,0} + \alpha_{0,u} U + \alpha_{0,x} X) \right \}$, $M \sim \cN(\tau_0 + \tau_a A + \tau_u U + \tau_x X, \sigma_{m|u,a,x}^2)$ given $U$, $A$ and $X$, and $Z \sim \cN(\epsilon_0 + \epsilon_u U + \epsilon_a A + \epsilon_x X, \sigma_{z|u, a, x}^2)$ given $U$, $A$ and $X$, then the choices of exposure confounding bridge functions:
    \begin{align*}
      q_0(Z,X) =& \exp\left \{ -(\gamma_{0,0} + \gamma_{0,z}Z + \gamma_{0,x}X) \right \}, \\
      q_1(Z,M,X) =& q_0(Z,X) \exp\left \{ \gamma_{1,0} + \gamma_{1,z}Z + \gamma_{1,m}M + \gamma_{1,x}X \right \},
    \end{align*}
    satisfies the integral equations~\eqref{eq: q0} and \eqref{eq: q1}, at the parameter values:
    \begin{equation*}
       \left\{\begin{matrix}
        \gamma_{0,0} = \alpha_{0,0} - \frac{\alpha_{0,u}}{\epsilon_u} \left( \epsilon_0 - \frac{\alpha_{0,u} \sigma_{z|u, a, x}^2}{2\epsilon_u} \right) \\
        \gamma_{0,z} = \frac{\alpha_{0,u}}{\epsilon_u} \\
         \gamma_{0,x} = \alpha_{0,x} - \frac{\alpha_{0,u} \epsilon_x}{\epsilon_u}
       \end{matrix}\right.
    \end{equation*}
    and
    \begin{equation*}
       \left\{\begin{matrix}
        \gamma_{1,0} = \alpha_{1,0} - \frac{\alpha_{1,u}}{\epsilon_u} \left( \epsilon_0 + \epsilon_a + \frac{\alpha_{1,u} \sigma_{z|u, a, x}^2}{2\epsilon_u} \right) \\
        \gamma_{1,m} = \alpha_{1,m} \\
        \gamma_{1,z} = \frac{\alpha_{1,u}}{\epsilon_u} \\
         \gamma_{1,x} = \alpha_{1,x} - \frac{\alpha_{1,u} \epsilon_x}{\epsilon_u}
       \end{matrix}\right.
    \end{equation*}
    where
    \begin{equation*}
       \left\{\begin{matrix}
         \alpha_{1,0} = \frac{\tau_a}{\sigma_{m|u,a,x}^2} \left( \frac{\tau_a}{2} + \tau_0 \right) + \alpha_{0,0} \\
         \alpha_{1,m} = -\frac{\tau_a}{\sigma_{m|u,a,x}^2} \\
         \alpha_{1,u} = \frac{\tau_a \tau_u}{\sigma_{m|u,a,x}^2} + \alpha_{0,u} \\
         \alpha_{1,x} = \frac{\tau_a \tau_x}{\sigma_{m|u,a,x}^2} + \alpha_{0,x}
      \end{matrix}\right.
    \end{equation*}
    are parameters of $\log \frac{f(A=0|U,M,X)}{f(A=1|U,M,X)} = \alpha_{1,0} + \alpha_{1,m} M + \alpha_{1,u} U + \alpha_{1,x} X$.
\end{corollary}

\newpage 
\bibliographystyle{apalike}
\bibliography{mediation}

\end{document}